\documentclass[a4paper,11pt]{article}
\pdfoutput=1 

\usepackage{jcappub} 

\usepackage[T1]{fontenc} 
\usepackage{graphics}
\usepackage{lscape}
\usepackage[left]{lineno}
\usepackage{booktabs} 

\usepackage{tabularx,multirow} 
\usepackage{rotating} 

\title{Time-dependent search for neutrino emission from X-ray binaries with the ANTARES telescope}

\usepackage{longtable}

\author[1]{A.~Albert}
\author[2]{M.~Andr\'e}
\author[3]{G.~Anton}
\author[4]{M.~Ardid}
\author[5]{J.-J.~Aubert}
\author[6]{T.~Avgitas}
\author[6]{B.~Baret}
\author[7]{J.~Barrios-Mart\'{\i}}
\author[8]{S.~Basa}
\author[5]{V.~Bertin}
% ...

\affiliation[1]{\scriptsize{GRPHE - Universit\'e de Haute Alsace - Institut universitaire de technologie de Colmar, 34 rue du Grillenbreit BP 50568 - 68008 Colmar, France}}
\affiliation[2]{\scriptsize{Technical University of Catalonia, Laboratory of Applied Bioacoustics, Rambla Exposici\'o,08800 Vilanova i la Geltr\'u,Barcelona, Spain}}
\affiliation[3]{\scriptsize{Friedrich-Alexander-Universit\"at Erlangen-N\"urnberg, Erlangen Centre for Astroparticle Physics, Erwin-Rommel-Str. 1, 91058 Erlangen, Germany}}
\affiliation[4]{\scriptsize{Institut d'Investigaci\'o per a la Gesti\'o Integrada de les Zones Costaneres (IGIC) - Universitat Polit\`ecnica de Val\`encia. C/  Paranimf 1 , 46730 Gandia, Spain.}}
\affiliation[5]{\scriptsize{Aix-Marseille Universit\'e, CNRS/IN2P3, CPPM UMR 7346, 13288 Marseille, France}}
\affiliation[6]{\scriptsize{APC, Universit\'e Paris Diderot, CNRS/IN2P3, CEA/IRFU, Observatoire de Paris, Sorbonne Paris Cit\'e, 75205 Paris, France}}
\affiliation[7]{\scriptsize{IFIC - Instituto de F\'isica Corpuscular (CSIC - Universitat de Val\`encia) c/ Catedr\'atico Jos\'e Beltr\'an, 2 E-46980 Paterna, Valencia, Spain}}
\affiliation[8]{\scriptsize{LAM - Laboratoire d'Astrophysique de Marseille, P\^ole de l'\'Etoile Site de Ch\^ateau-Gombert, rue Fr\'ed\'eric Joliot-Curie 38,  13388 Marseille Cedex 13, France}}
% ...

\emailAdd{dornic@cppm.in2p3.fr}
\emailAdd{agustin.sanchez@ba.infn.it}
\emailAdd{coleiro@apc.univ-paris7.fr}

\abstract{ANTARES is currently the largest neutrino telescope operating in the Northern Hemisphere, aiming at the detection of 
high-energy neutrinos from astrophysical sources. Neutrino telescopes constantly monitor at least one complete 
hemisphere of the sky, and are thus well-suited to detect neutrinos produced in transient astrophysical sources. A time-dependent 
search has been applied to a list of 33~X-ray binaries undergoing high flaring activities in satellite data (RXTE/ASM, MAXI and Swift/BAT) 
and during hardness transition states in the~2008--2012 period. The background 
originating from interactions of charged cosmic rays in the Earth's atmosphere is drastically reduced by requiring a 
directional and temporal coincidence with astrophysical phenomena. The results of this search are presented together with 
comparisons between the neutrino flux upper limits and the neutrino flux predictions from astrophysical models. The neutrino flux upper limits
resulting from this search limit the jet parameter space for some astrophysical models.}

\begin{document}
\maketitle
\flushbottom

\section{Introduction}
X-ray binaries are binary systems composed of a compact object (neutron star (NS)
or stellar mass black hole (BH) candidate) and a companion non-degenerate star. Due to the strong gravitational attraction, matter expelled from the companion is accreted by the compact 
object. Depending on the mass of the companion star and the process of matter accretion, X-ray binaries are separated into two classes: 
Low-Mass X-ray Binaries (LMXB) which contain an evolved companion star of spectral class later than B transferring matter to the compact 
object through Roche lobe overflows; and High-Mass X-ray Binaries (HMXB) consisting of a massive O or B star developing intense stellar 
winds, a fraction of which is accreted by the compact object. While some of these objects are seen as persistent sources, most of them 
exhibit occasional outbursts, making them transient sources, in particular in the radio and X-ray domains.

Recent detections of GeV--TeV gamma-ray signals from some X-ray binaries confirm that they can produce outflows containing particles 
accelerated away from the compact object up to relativistic speeds~\cite{bib:Tavani2009}. At the moment, it is not clear whether the 
high-energy particle acceleration is a common process occurring in X-ray binaries but observed only in some systems with preferred 
(geometrical) characteristics with respect to the line of sight, or whether it is powered by a different mechanism at work only in 
some specific systems.

The theoretical mechanisms of gamma-ray production from X-ray binaries generally assume (very-) high-energy photon emission 
from the interaction of a relativistic outflow from the compact object with the wind and radiation emitted by the companion star. The 
outflow can take different shapes. In microquasars~\cite{bib:MirabelRodriguez1994} the high-energy emission is due to accretion energy 
released in the form of a collimated relativistic jets, detected in the radio domain through synchrotron emission. On the contrary, in 
other binary systems, high-energy emission can occur in a wide-angle shocked region, at the interface between pulsar and stellar 
winds~\cite{bib:Dubus2013}. They are probably the sites of effective acceleration of particles (electrons and/or protons) to multi-TeV energies but the nature of the high-energy emission is still unknown, and leptonic or hadronic origin is still debated nowadays~\cite{bib:Vila2010,bib:Vila2013, bib:Pepe2015}. 

Even if a rich variety of binary systems seems to be cosmic accelerators, some major issues are still open: are jets a common feature 
of X-ray binary systems? What is the particle acceleration mechanism at work in these systems? Is it unique? Constraining the jet 
composition and its baryonic content will help answering these questions. Indeed, the jet composition should be affected by the 
outflow-launching processes. For instance, jets powered by an accretion disk are likely to contain baryons~\cite{bib:BlandfordPayne1982} 
while jets which get their power from black hole spin are expected to be purely leptonic~\cite{bib:DiazTrigo2013}. Up to now, a hadronic 
component has been identified in only two X-ray binaries (SS 433 and 4U 1630-472)~\cite{bib:Migliari2002,bib:DiazTrigo2013} while a 
population of cold baryons present in the relativistic jet of Cyg\,X\texttt{-}1 has been proposed~\cite{bib:Heinz2006}.

Hadronic models of jet interactions with the winds
of massive stars were developed these last decades. The dominant hadronic contributions are expected from the 
photo-hadronic (p-$\gamma$) interactions between relativistic protons and synchrotron photons in the jet or 
coming from external sources ~\cite{bib:Levinson2001, bib:Distefano2002}), and from the hadronic (p-p) interactions 
between relativistic protons from the jet and thermal protons from the stellar 
wind~\cite{bib:Romero2003, bib:Christiansen2006,bib:Torres2007,bib:Vieyro2012}. In the absence of a jet, neutrinos can 
be produced through p-p processes between the accelerated protons in the the rotation-driven relativistic wind from the 
young neutron star and the circumstellar disk in the case of Be type stars~\cite{bib:NeronovRibordy2009,bib:Sahakyan2014}. 
The detection of high-energy neutrinos from an X-ray binary system would definitively confirm the presence of relativistic 
protons in the outflow, and thus further constrain the particle acceleration mechanism.

The ANTARES Collaboration completed the construction of a neutrino telescope in the Mediterranean Sea with the connection of its 
twelfth detector line in May 2008~\cite{bib:Antares}. The telescope is located 20 km off the Southern coast of France 
(42$^\circ$48'N, 6$^\circ$10'E), at a depth of 2475 m. In the ANTARES telescope, events are primarily detected by 
observing the Cherenkov light induced by relativistic muons in the darkness of the deep sea. Owing to their low interaction 
probablility, only neutrinos have the ability to cross the Earth. Therefore, an upgoing muon is an unambiguous signature of a 
neutrino interaction close to the detector. To distinguish astrophysical neutrino events from background events (muons and neutrinos) 
generated in the atmosphere, energy and direction reconstructions have been used in several searches~\cite{bib:PointSource,bib:Diffuse}. 
To improve the signal-to-noise discrimination, the arrival time information can be used to significantly reduce the effective 
background~\cite{bib:MDP}. 

In this paper, the results of a time-dependent search for cosmic neutrino sources using the ANTARES data taken from~2008 to~2012 is 
presented. This extends a previous ANTARES analysis~\cite{bib:AntaresMicroQ} where only six sources and the first three years of 
data-taking were considered. It is also complementary to a previous IceCube transient analysis~\cite{bib:IceCube2015} which 
considered few X-ray binary systems. However, the ANTARES location in the northern hemisphere, and its lower neutrino energy threshold in 
comparison with IceCube, make it well-suited to study neutrino emission from such galactic sources. Neutrino emission has been 
searched-for during outburst periods of X-ray binaries characterised by the variability of their soft and hard X-ray flux 
density~\cite{bib:Remillard}. Jet emission, probably linked to particle acceleration and thus potential neutrino emission, usually 
occurs during periods of high levels of hard X-ray flux density (called hard states) and during transition periods (intermediate states) 
between a hard state and a soft state. Sections~2 and~3 present the selection of outburst periods selection from X-ray light curves, 
during hard and intermediate states respectively. Section~4 details the statistical approach 
used to perform the analysis, while results are provided and discussed in Section~5. Conclusions are drawn in Section~6.

\section{Selection of outburst periods}

The time-dependent analysis described in the following sections is applied to a list of X-ray binaries exhibiting outburst periods in their light curves. In the following, particle acceleration is assumed to take place in the XRB and neutrino emission is assumed to be correlated with hard X-ray outbursts. This is generally the case for microquasars exhibiting relativistic jets. Indeed, a correlation is expected between hard X-ray outbursts (usually produced by both synchrotron and inverse Compton mechanisms) and compact jet emission where cosmic ray acceleration may occur~\cite{bib:Corbel2001}. It can be noticed that larger Lorentz factors and potentially larger neutrino emission might be expected during transition states, when transient outbursts are observed. These specific states of microquasars will be treated in Section 3.

However, most of the X-ray binaries included in our sample are not classified as microquasars. This might not prevent neutrino emission, in particular during hard X-ray outburst phases.
For instance, the LMXB GRO J0422+32 does not present any clear evidence of a relativistic jet although hard X-ray flares are observed \cite{bib:MillerJones2011}. During these outbursts, cosmic rays might be accelerated through magnetic reconnection in the corona (where hard X-ray photons are produced) surrounding the black hole, as suggested by \cite{bib:Vieyro2012a}. X-ray binaries containing a magnetized NS do not usually exhibit relativistic ejection of matter. However, as observed in systems such as Cir X-1~\cite{bib:MillerJones2012}, transient jets can be produced upon a sufficiently low ($\lesssim 10^8$ G) magnetic field magnitude at the surface of the NS~\cite{bib:Sguera2009}, based on a similar accretion/ejection mechanism as microquasars. In supergiant fast X-ray transients (HMXB composed of a NS and exhibiting rapid hard X-ray flux evolution) magnetic field decay could also allow for relativistic particle injection when large clumps of matter are accreted by the NS~\cite{bib:Garcia2014}, producing again a potential correlation between hard X-ray flares and particle acceleration.

Finally, when the magnetosphere of an accreting NS penetrates the decretion disk of the compagnon star (as observed in HMXB composed of a Be type companion star), intense hard X-ray emission is expected \cite{bib:Giovannelli1992}. Meanwhile, in the case of magnetized NS, the magnetosphere may develop electrostatic gaps where protons could be accelerated along the magnetic field lines, thus allowing for neutrino production~\cite{bib:Anchordoqui2003}.\\

\begin{figure}[ht!]
\centering
\includegraphics[width=\textwidth]{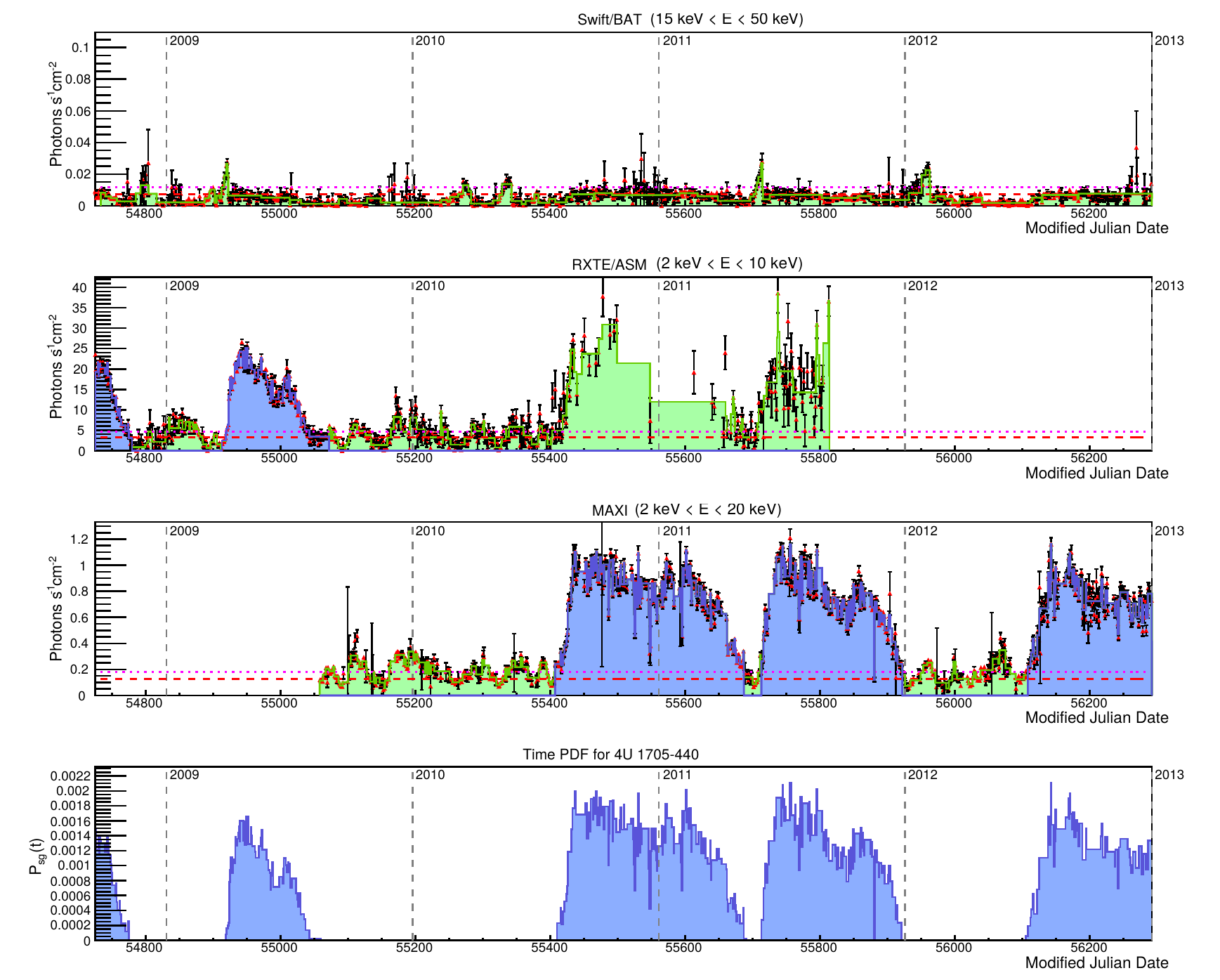}
\caption{Light curves for 4U 1705-440 as seen by Swift/BAT, RXTE/ASM and MAXI during the studied period. The estimated baseline 
emissions, $BL$, (red lines) and $BL + 1\sigma_{BL}$ (magenta lines) are also shown. Histograms correspond to the light curves treated 
with the MLB algorithm. The blue histograms represent the selected flaring periods of each light curve, merged to produce the time PDF (bottom).
Green histograms show periods of each light curve not selected for this analysis.}
\label{fig:lcsample}
\end{figure}

Light curves used for the selection of outburst periods are obtained mainly using the Swift/BAT telescope\footnotemark[1]. These data are complemented by those 
from other instruments: RXTE/ASM\footnotemark[2] and MAXI\footnotemark[3]. A maximum likelihood block (MLB) algorithm~\cite{bib:Scargle} 
is used to remove noise from the light curve by iterating over the data points and selecting periods during which data are consistent 
with a constant flux within statistical errors. This algorithm is applied independently to all the light curves from all the satellites. 
Depending on the time period and the availability of the different instruments, outbursts are better observed in one apparatus compared 
to others. As the energy range and the sensitivity of these telescopes are different, it is not easy to combine measurements into a 
single time-dependent function to describe the light curve. The value of the steady state (i.e. baseline, $BL$) and its fluctuation 
($\sigma_{BL}$) are determined with a Gaussian fit of the lower part of the distribution of the flux. The baseline is 
subtracted from the light curve and the amplitude is converted to a relative amplitude by dividing by $\sigma_{BL}$. 
Finally, the relative light curves from different instruments are merged, as can be seen for the sample source 4U~1705-440 in 
Figure~\ref{fig:lcsample}. This is used to produce the time probability functions for the analysis (see Section 4).

\footnotetext[1]{http://swift.gsfc.nasa.gov/results/transients}
\footnotetext[2]{http://xte.mit.edu/ASM\_lc.html}
\footnotetext[3]{http://maxi.riken.jp}

The flaring periods are defined from the blocks of the light curve characterised by the MLB algorithm in three main steps. Firstly, 
seeds are identified by searching for blocks with an amplitude above $BL\,+\,8\sigma_{BL}$. Then, each period is extended forward and 
backward up to an emission compatible with $BL\,+\,1\sigma_{BL}$. A delay of 0.5~days is added before and after the flare in order to 
take into account that the precise time of the flare is not known (one-day binned light curve)  and the potential delay between the X-ray and neutrino emissions. Finally, spurious flares are discarded 
if they are not visible by at least one other instrument. The final list includes 33~X-ray binaries: 1~HMXB (BH), 11~HMXB (NS), 8~HMXB 
(BH candidate), 10~LMXB (NS) and 3~XRB (BH candidate), as reported in Table~\ref{table:Sources}. Table \ref{table:Outbursts} summarizes for each source the Modified Julian Date of 
 start-stop of the flare, and the satellite that provided the information.

\begin{table}[ht!]
\caption{List of 33 X-ray binaries with significant flares selected for this analysis.}
\label{table:Sources}
\centering
\begin{tabular}{|c||c|c|c|}
\hline Name    & Class & {RA [$^\circ$]} & {Dec [$^\circ$]} \\
\hline
\hline{Cyg\,X\texttt{-}1}            & HMXB (BH)  & 230.170 & -57.167 \\
\hline{1A\,0535\texttt{+}262}        & HMXB (NS)  & ~84.727 & ~26.316 \\
\hline{1A\,1118\texttt{-}61}         & HMXB (NS)  & 170.238 & -61.917 \\
\hline{Ginga\,1843\texttt{+}00}      & HMXB (NS)  & 281.404 & ~~0.863 \\
\hline{GS\,0834\texttt{-}430}        & HMXB (NS)  & 128.979 & -43.185 \\
\hline{GX\,304\texttt{-}1}           & HMXB (NS)  & 195.321 & -61.602 \\
\hline{H\,1417\texttt{-}624}         & HMXB (NS)  & 215.303 & -62.698 \\
\hline{MXB\,0656\texttt{-}072}       & HMXB (NS)  & 104.572 & ~-7.210 \\
\hline{XTE\,J1946\texttt{+}274}      & HMXB (NS)  & 296.414 & ~27.365 \\
\hline{GX\,1\texttt{+}4}             & HMXB (NS)  & 263.009 & -24.746 \\
\hline{MAXI\,J1409\texttt{-}619}     & HMXB (NS)  & 212.011 & -61.984 \\
\hline{GRO\,J1008\texttt{-}57}       & HMXB (NS)  & 152.433 & -58.295 \\
\hline{GX\,339\texttt{-}4}           & LMXB (BHC) & 255.706 & -48.784 \\
\hline{4U\,1630\texttt{-}472}        & LMXB (BHC) & 248.504 & -47.393 \\
\hline{IGR\,J17091\texttt{-}3624}    & LMXB (BHC) & 257.282 & -36.407 \\
\hline{IGR\,J17464\texttt{-}3213}    & LMXB (BHC) & 266.565 & -32.234 \\
\hline{MAXI\,J1659\texttt{-}152}     & LMXB (BHC) & 254.757 & -15.258 \\
\hline{SWIFT\,J1910.2\texttt{-}0546} & LMXB (BHC) & 287.595 & ~-5.799 \\
\hline{XTE\,J1752\texttt{-}223}      & LMXB (BHC) & 268.063 & -22.342 \\
\hline{SWIFT\,J1539.2\texttt{-}6227} & LMXB (BHC) & 234.800 & -62.467 \\
\hline{4U\,1954\texttt{+}31}         & LMXB (NS)  & 298.926 & ~32.097 \\
\hline{Aql\,X\texttt{-}1}            & LMXB (NS)  & 287.817 & ~~0.585 \\
\hline{Cir\,X\texttt{-}1}            & LMXB (NS)  & 230.170 & -57.167 \\
\hline{EX\,O1745\texttt{-}248}       & LMXB (NS)  & 267.022 & -24.780 \\
\hline{H\,1608\texttt{-}522}         & LMXB (NS)  & 243.179 & -52.423 \\
\hline{SAX\,J1808.4\texttt{-}3658}   & LMXB (NS)  & 272.115 & -36.977 \\
\hline{XTE\,J1810\texttt{-}189}      & LMXB (NS)  & 272.586 & -19.070 \\
\hline{4U\,1636\texttt{-}536}        & LMXB (NS)  & 250.231 & -53.751 \\
\hline{4U\,1705\texttt{-}440}        & LMXB (NS)  & 257.225 & -44.102 \\
\hline{IGR\,J17473\texttt{-}2721}    & LMXB (NS)  & 266.825 & -27.344 \\
\hline{MAXI\,J1836\texttt{-}194}     & XRB (BHC)  & 278.931 & -19.320 \\
\hline{XTE\,J1652\texttt{-}453}      & XRB (BHC)  & 253.085 & -45.344 \\
\hline{SWIFT\,J1842.5\texttt{-}1124} & XRB (BHC)  & 280.573 & -11.418 \\
\hline
\end{tabular}
\end{table}

\begin{table}[ht!]
\caption{List of the studied flaring periods and the satellite used (R: Rossi; S: Swift; M: MAXI).}
\label{table:Outbursts}
\centering
\begin{footnotesize}

\begin{tabular}{|c||c|}

\hline Name    & Flaring periods (MJD) \\
\hline
\hline{\multirow{1}{*}{Cyg\,X\texttt{-}1}}            & 54715--55468 (S), 55662--55804 (S), 55888--55947 (S), 56011--56098 (S) \\
\hline{\multirow{3}{*}{1A\,0535\texttt{+}262}}        & 54719--54731 (S), 54834--54845 (S), 54936--54954 (S), 55034--55070 (S), 55160--55204 (S) \\
                                                      & 55272--55313 (S), 55370--55425 (S), 55479--55503 (S), 55602--55647 (S) \\
\hline{1A\,1118\texttt{-}61}                          & 54827--54907 (S) \\
\hline{Ginga\,1843\texttt{+}00}                       & 54897--54928 (S) \\
\hline{GS\,0834\texttt{-}430}                         & 56095--56180 (S) \\
\hline{\multirow{2}{*}{GX\,304\texttt{-}1}}           & 55146--55165 (S), 55276--55301 (S), 55411--55439 (S), 55534--55566 (S), 55672--55703 (S) \\
                                                      & 55802--55835 (S), 55932--55973 (S), 56062--56119 (S), 56186--56203 (S) \\
\hline{H\,1417\texttt{-}624}                          & 55116--55222 (S) \\
\hline{MXB\,0656\texttt{-}072}                        & 54724--54762 (S) \\
\hline{\multirow{1}{*}{XTE\,J1946\texttt{+}274}}      & 55351--55411 (S), 55442--55479 (S), 55510--55561 (S), 55575--555621 (S), 55640--55688 (S) \\
\hline{\multirow{2}{*}{GX\,1\texttt{+}4}}             & 54800--54854 (R), 54948--54953 (S), 54970--55088 (S), 55113--55206 (S) \\
                                                      & 55223--55264 (S), 55425--55496 (S), 55509--55580 (S), 55619--55652 (S), 55707--55778 (S) \\
\hline{MAXI\,J1409\texttt{-}619}                      & 55487--55523 (S) \\
\hline{\multirow{2}{*}{GRO\,J1008\texttt{-}57}}       & 54910--54928 (S), 55156--55175 (S), 55408--55427 (S) \\
                                                      & 55659--55678 (S), 55905--55929 (S), 56152--56291 (S) \\
\hline{\multirow{1}{*}{GX\,339\texttt{-}4}}           & 54880--55014 (S), 55183--55337 (S), 55557--55601 (S) \\
\hline{\multirow{1}{*}{4U\,1630\texttt{-}472}}        & 55193--55285 (R), 55301--55331 (R), 55919--55930 (S), 56041--56106 (S), 56176--56221 (S) \\
\hline{IGR\,J17091\texttt{-}3624}                     & 55588--55649 (S) \\
\hline{\multirow{2}{*}{IGR\,J17464\texttt{-}3213}}    & 54732--54803 (S), 54972--55036 (S), 55175--55191 (S), 55411--55468 (S) \\
                                                      & 55657--55711 (S), 55912--55972 (S), 56192--56234 (S) \\
\hline{MAXI\,J1659\texttt{-}152}                      & 55464--55525 (S) \\
\hline{\multirow{1}{*}{SWIFT\,J1910.2\texttt{-}0546}} & 55946--55951 (M), 56093--56181 (S), 56211--56291 (S) \\
\hline{XTE\,J1752\texttt{-}223}                       & 55127--55367 (S) \\
\hline{SWIFT\,J1539.2\texttt{-}6227}                  & 54793--54838 (S) \\
\hline{\multirow{2}{*}{4U\,1954\texttt{+}31}}         & 54772--54791 (S), 55246--55257 (S), 55363--55400 (S) \\
                                                      & 56159--56171 (S), 56189--56199 (S), 56288--56291 (S) \\
\hline{\multirow{1}{*}{Aql\,X\texttt{-}1}}            & 54901--54922 (S), 55140--55276 (S), 55390--55485 (S), 55844--55911 (S) \\
\hline{\multirow{2}{*}{Cir\,X\texttt{-}1}}            & 55316--55385 (M), 55576--55649 (M), 55743--55777 (M), 55778--55803 (M) \\
                                                      & 55840--55876 (M), 55895--55951 (M), 56150--56173 (M) \\
\hline{\multirow{1}{*}{EX\,O1745\texttt{-}248}}       & 55477--55541 (S), 56119--56145 (S) \\
\hline{\multirow{2}{*}{H\,1608\texttt{-}522}}         & 54715--54786 (S), 54860--54894 (S), 54900--54921 (S), 54930--54978 (S) \\
                                                      & 54998--55089 (S), 55257--55281 (M), 55643--55993 (S), 56208--56253 (M) \\
\hline{\multirow{1}{*}{SAX\,J1808.4\texttt{-}3658}}   & 54731--54746 (S), 55866--55885 (S) \\
\hline{XTE\,J1810\texttt{-}189}                       & 56271--56291 (S) \\
\hline{4U\,1636\texttt{-}536}                         & 56114--56160 (S) \\
\hline{\multirow{2}{*}{4U\,1705\texttt{-}440}}        & 54715--54780 (R), 54919--55060 (R), 55410--55691 (M) \\
                                                      & 55713--55921 (M), 56105--56291 (M) \\
\hline{IGR\,J17473\texttt{-}2721}                     & 54715--54734 (S) \\
\hline{\multirow{1}{*}{MAXI\,J1836\texttt{-}194}}     & 55222--55273 (R), 55802--55884 (S) \\
\hline{XTE\,J1652\texttt{-}453}                       & 55011--55076 (S) \\
\hline{SWIFT\,J1842.5\texttt{-}1124}                  & 54715--54858 (S) \\
\hline

\end{tabular}
\end{footnotesize}

\end{table}

\section{Selection of transition state periods}\label{TSperiods}

Spectral transition states of XRB are difficult to define: there are no regular observations with X-ray satellites, and statistics are very low due to the inaccurate measurement of the hardness ratio, defined as the ratio of counts in different X-ray wavebands. Considering the difficulties of an extensive coverage of the transition states of XRB, the selection of the these transition periods relies on the alerts reported in \textit{Astronomer's Telegram}~\footnotemark[4]. The selected 19 alerts in the period 2008--2012, distributed among 8 sources, and the corresponding transition periods, are shown in Table~\ref{table:SourcesHardness}. Considering the lack of information on the time variation of flux, a constant emission during each of the reported dates has been assumed.


\footnotetext[4]{$http://www.astronomerstelegram.org$}

\begin{table}[ht!]
\caption{List of 8 X-ray binaries with hardness transition states reported in \textit{Astronomer's Telegram}.}
\label{table:SourcesHardness}
\centering
\begin{tabular}{|c||c|c|}
\hline Name    & \texttt{\#ATel} & {Transition State Periods [MJD] (days)} \\
\hline
\hline{GX\,339\texttt{-}4}           & \multicolumn{1}{c|}{\begin{tabular}[x]{@{}c@{}}\texttt{\#2577}~~\texttt{\#2593}\\\texttt{\#3117}~~\texttt{\#3191}\end{tabular}} & \multicolumn{1}{c|}{\begin{tabular}[x]{@{}c@{}}$55303$\,--\,$55305\;(\;2\;)$~~~~$55308$\,--\,$55309\;(\;1\;)$\\$55315$\,--\,$55316\;(\;1\;)$~~~~$55318$\,--\,$55319\;(\;1\;)$\\$55580$\,--\,$55581\;(\;1\;)$~~~~$55616$\,--\,$55617\;(\;1\;)$\end{tabular}} \\
\hline{H\,1608\texttt{-}522}         & \texttt{\#2072}~~\texttt{\#2467} & \multicolumn{1}{c|}{\begin{tabular}[x]{@{}c@{}}$54960$\,--\,$54976\;(16)$\end{tabular}} \\
\hline{IGR\,J17091\texttt{-}3624}    & \texttt{\#3179}~~\texttt{\#3196} & \multicolumn{1}{c|}{\begin{tabular}[x]{@{}c@{}}$55611$\,--\,$55612\;(\;1\;)$~~~~$55962$\,--\,$55964\;(\;2\;)$\end{tabular}} \\
\hline{IGR\,J17464\texttt{-}3213}    & \multicolumn{1}{c|}{\begin{tabular}[x]{@{}c@{}}\texttt{\#1804}~~\texttt{\#1813}\\\texttt{\#3301}~~\texttt{\#3842}\end{tabular}} & \multicolumn{1}{c|}{\begin{tabular}[x]{@{}c@{}}$54752$\,--\,$54759\;(\;7\;)$~~~~$55671$\,--\,$55672\;(\;1\;)$\\$55925$\,--\,$55927\;(\;2\;)$\end{tabular}} \\
\hline{MAXI\,J1659\texttt{-}152}     & \texttt{\#2951}~~\texttt{\#2999} & \multicolumn{1}{c|}{\begin{tabular}[x]{@{}c@{}}$55481$\,--\,$55487\;(\;6\;)$~~~~$55500$\,--\,$55502\;(\;2\;)$\end{tabular}} \\
\hline{SWIFT\,J1910.2\texttt{-}0546} & \texttt{\#4139}~~\texttt{\#4273} & \multicolumn{1}{c|}{\begin{tabular}[x]{@{}c@{}}$56094$\,--\,$56095\;(\;1\;)$~~~~$56131$\,--\,$56133\;(\;2\;)$\end{tabular}} \\
\hline{XTE\,J1652\texttt{-}453}      & \texttt{\#2219} & \multicolumn{1}{c|}{\begin{tabular}[x]{@{}c@{}}$55010$\,--\,$55085\;(75)$\end{tabular}} \\
\hline{XTE\,J1752\texttt{-}223}      & \texttt{\#2391}~~\texttt{\#2518} & \multicolumn{1}{c|}{\begin{tabular}[x]{@{}c@{}}$55219$\,--\,$55220\;(\;1\;)$~~~~$55492$\,--\,$55493\;(\;1\;)$\end{tabular}} \\
\hline
\end{tabular}
\end{table}

\section{Time-dependent analysis}

The ANTARES data collected between 2008 and 2012, corresponding to 1044 days of lifetime, are analysed to search for neutrino events around 
the selected sources, in coincidence with the time periods defined in the previous sections. The statistical method adopted to infer the 
presence of a signal on top of the atmospheric neutrino background, or alternatively set upper limits on the neutrino flux is an unbinned 
method based on an extended maximum likelihood ratio test statistic. It has been previously used to search for neutrinos from gamma-ray flaring 
blazars~\cite{bib:AntBlazar}. The likelihood, $\mathcal{L}$, is defined as:
\begin{equation}
\ln \mathcal{L(\mathcal{N}_{\rm S})} = \left(\sum_{i=1}^{N} \ln[\mathcal{N}_{\rm S}\mathcal{S}_{i}+\mathcal{N}_{\rm B}\mathcal{B}_{i}]\right)-[\mathcal{N}_{\rm S}+\mathcal{N}_{\rm B}]
\label{eq:EQ_likelihood2}
\end{equation}
\noindent where $\mathcal{S}_{i}$ and $\mathcal{B}_{i}$ are the probabilities for signal and background for an event $i$, respectively, 
and $\mathcal{N}_{\rm S}$ (not known) and $\mathcal{N}_{\rm B}$ (known) are the number of expected signal and background events in the data 
sample. $N$ is the total number of events in the considered data sample. To discriminate the signal-like events from background, these probabilities are described by the product of three components 
related to the direction, energy, and timing of each event. For an event \textit{i}, the signal probability is:
\begin{equation}
\mathcal{S}_{i} = \mathcal{S}^{\rm space}(\Psi_{i})\cdot \mathcal{S}^{\rm energy}(dE/dX_{i})\cdot \mathcal{S}^{\rm time}(t_{i}+t_{\rm lag})
\label{eq:EQ_likelihood3}
\end{equation}

Here, $\mathcal{S}^{\rm space}$ is a parametrisation of the point spread function, i.e., $\mathcal{S}^{\rm space}(\Psi_{i})$ is the 
probability to reconstruct an event \textit{i} at an angular distance $\Psi_{i}$ from the true source location. The energy PDF, 
$\mathcal{S}^{\rm energy}$, is the normalised distribution of the muon energy estimator, dE/dX, of an event according 
to the studied energy spectrum. To cover the majority of the range allowed by the models~\cite{bib:Distefano2002,bib:Kappes} accessible to 
the ANTARES sensitivity, three neutrino-energy differential spectra are tested in this analysis: $E^{-2}$, $E^{-2}\exp(-E/100~\rm{TeV})$ 
and $E^{-2}\exp(-E/10~\rm{TeV})$, where $E$ is the neutrino energy. The shape of the time PDF, $\mathcal{S}^{\rm time}$, for the signal 
event is extracted directly from the gamma-ray light curve parametrisation, as described in the previous sections, assuming 
proportionality between gamma-ray and neutrino fluxes. A possible lag of up to $\pm$5 days has been introduced in the likelihood to allow 
for small lags in the proportionality. This corresponds to a possible shift of the entire time PDF. The lag parameter is fitted in the 
likelihood maximisation together with the number of fitted signal events in the data. The background probability for an event \textit{i} is:
\begin{equation}
\mathcal{B}_{i} = \mathcal{B}^{\rm space}(\delta_{i})\cdot \mathcal{B}^{\rm energy}(dE/dX_{i})\cdot \mathcal{B}^{\rm time}(t_{i})
\label{eq:EQ_likelihood4}
\end{equation}
\noindent where the directional PDF, $\mathcal{B}^{\rm{space}}$, the energy PDF, $\mathcal{B}^{\rm{energy}}$, and the time 
PDF, $\mathcal{B}^{\rm{time}}$, for the background are derived from data using, respectively, the observed declination, $\delta_i$, 
distribution of selected events in the sample, the measured distribution of the energy estimator, and the observed time distribution of 
all the reconstructed muons.

The goal of the unbinned search is to determine, in a given direction in the sky and at a given time, the relative contribution of each 
component, and to calculate the probability to have a signal above a given background model. This is done via the test statistic, 
$\lambda$, defined as the ratio of the probability for the hypothesis of background plus signal over the probability 
of only background:
\begin{equation}
\lambda=\sum_{i=1}^{N} \ln\frac{\mathcal{L(\mathcal{N}_{\rm S})}}{\mathcal{L(\mathcal{N}_{\rm S}}=0)} 
\label{eq:TS}
\end{equation}

The evaluation of the test statistic is performed by generating pseudo-experiments simulating background and signal in a 30$^{\circ}$ cone 
around the considered source according to the background-only and background plus signal hypotheses. The performance of the time-dependent 
analysis is computed using toy experiments. For time ranges characteristic of flaring activity, the time-dependent search presented here improves 
the discovery potential by on-average a factor 2-3 with respect to a standard time-integrated point-source search~\cite{bib:PointSource}, 
under the assumption that the neutrino emission is correlated with the X-ray flaring activity.

\section{Results and discussions}
\subsection{Results}
\begin{figure}[ht!]
\centering
\includegraphics[width=\textwidth]{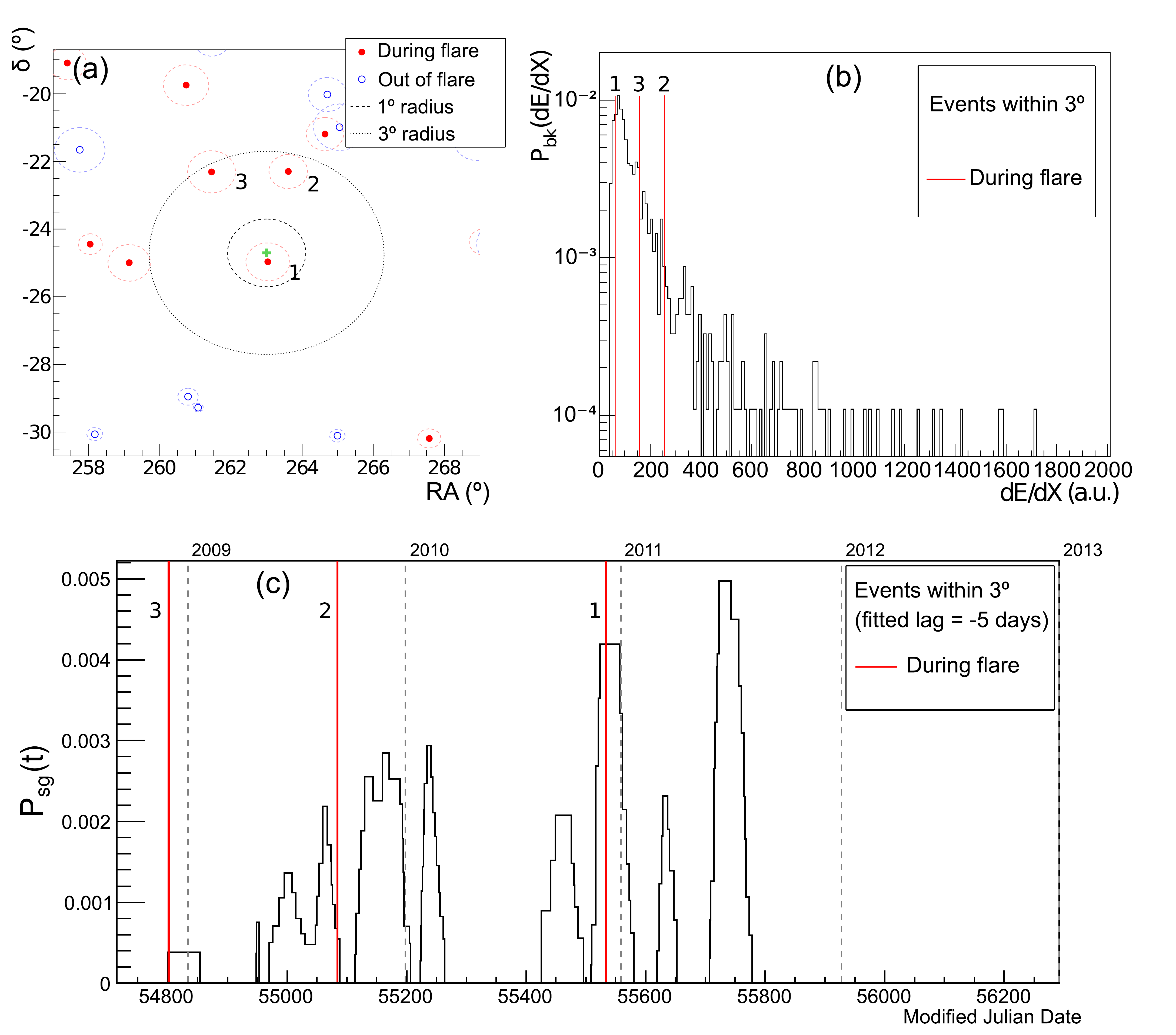}
\caption{Results of this analysis for GX\,1\texttt{+}4. (a) Event map around the direction of GX\,1\texttt{+}4 indicated by the green cross. 
The full red (hollow blue) dots indicate the events (not) in time coincidence with the selected flares. The size of the circle around the dots 
is proportional to the estimated angular uncertainty for each event. The three closest events from the source direction are labeled 1, 2 and 
3. (b) Distribution of the estimated energy dE/dX in a $\pm10^{\circ}$ declination band around the source direction. The red line displays the 
values of the three most significant events. (c) Time distribution of $\mathcal{B}^{\rm{time}}$. The red 
line displays the times of the 3 ANTARES events indicated in panel (a).}
\label{fig:GX14results}
\end{figure}

Only one source exhibited a significant signal excess during an X-ray flare: GX\,1\texttt{+}4, with 287 days of flare duration included in the 
analysis, shows a p-value of 4.1$\%$ with a fitted signal of 0.7~events and a lag of $-$4~days, which is obtained with the 
100 TeV cutoff energy spectrum. This result is due to one (three) events in a cone of 1~(3)~degrees in coincidence with X-ray outbursts 
detected by RXTE/ASM and Swift/BAT. Figure~\ref{fig:GX14results} shows the light curve of GX\,1\texttt{+}4 with the time of the neutrino 
candidates, the estimated energy distribution, and the angular distribution of the events around the position of this source. The post-trial 
probability, computed by taking into account the 33 searches, is 72$\%$, and is thus compatible with background fluctuations. In the hardness 
transition state analysis, no significant excess has been found, with a 77$\%$ post-trial probability for the full analysis.

\begin{figure}[ht!]
\centering
\includegraphics[width=\textwidth]{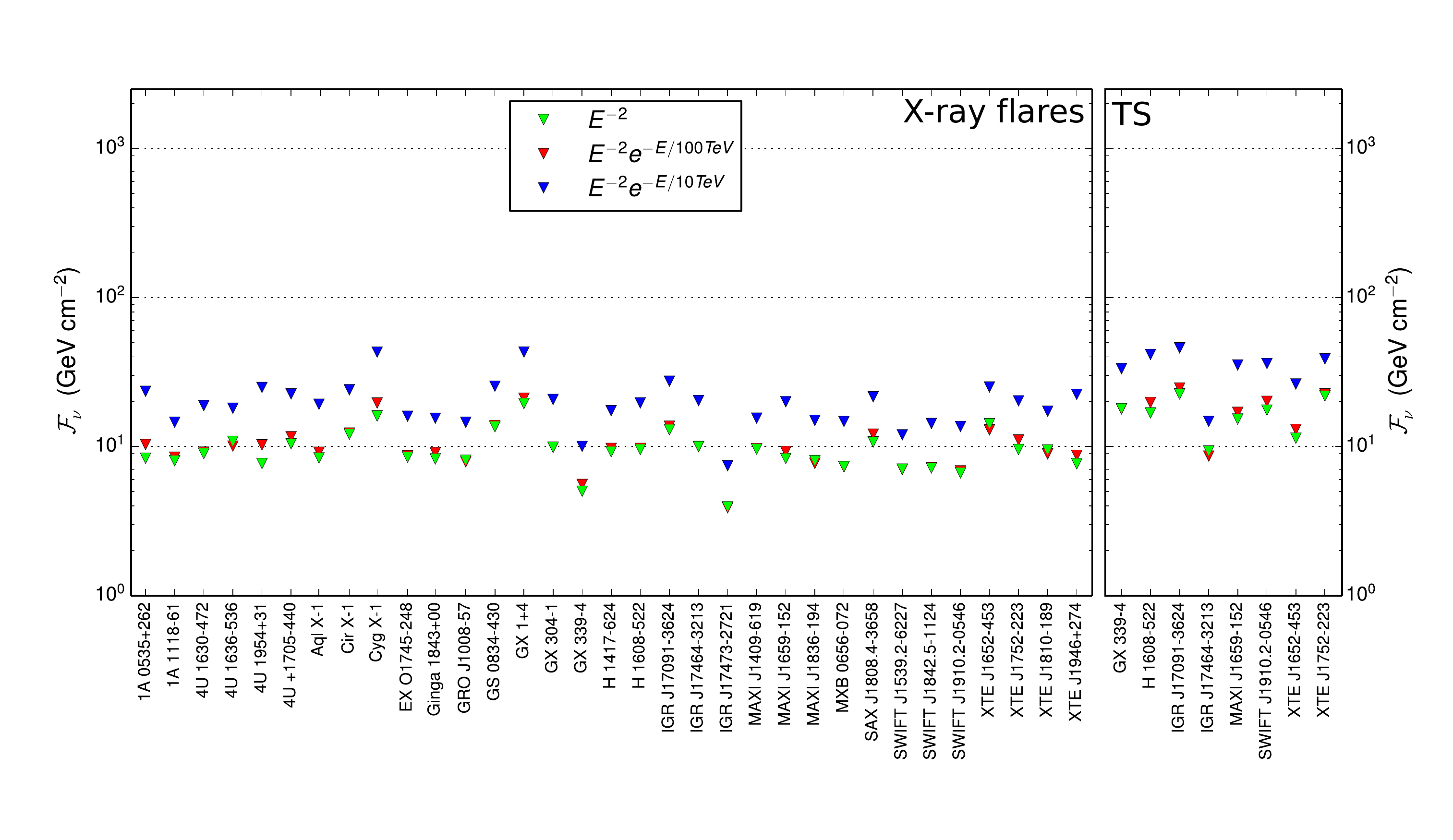}
\caption{Upper limits at 90\% C.L. on the neutrino fluence for the 33 XRB with outburst periods (left) and for the 8 XRB with transition state periods (right) in 
the case of $E^{-2}$ (green triangles), $E^{-2}\exp(-E/100~\rm{TeV})$ (red triangles) and $E^{-2}\exp(-E/10~\rm{TeV})$ (blue triangles) neutrino energy spectra. 
}
\label{fig:limit}
\end{figure}

In the absence of a discovery, upper limits on the neutrino fluence, $\mathcal{F}_{\nu}$, the energy flux, $F$, and the differential flux 
normalisation, $\phi_{0}$, at 90\% confidence level are computed using 5-95\% of the energy range as:
\begin{equation}
  \mathcal{F_\nu} = \Delta t \cdot F = \Delta t \int_{E_{\rm min}}^{E_{\rm max}} \mathrm{d}E \cdot E \cdot \phi_{0} \cdot \mathcal{S}(E) = \Delta t \cdot \phi_{0} \cdot \text{I}(E)
\end{equation}
\noindent where $\mathcal{S}(E)$ is the dimensionless neutrino spectrum, e.g. for the $E^{-2}$ spectrum$, 
\mathrm{d}N/\mathrm{d}E = \phi_{0} \cdot \mathcal{S}(E) = \phi_{0} \cdot (E/\text{GeV})^{-2}$. The limits are calculated according to the classical (frequentist) method for upper 
limits~\cite{bib:Neyman1937} and are summarised in Table~\ref{table:ULs}. Figure~\ref{fig:limit} displays these upper limits. Systematic uncertainties of 15\% on the angular 
resolution and 15\% on the detector acceptance have been included in the upper limit calculations~\cite{bib:PointSource}.\\

{
\setlength{\tabcolsep}{2pt}
\begin{sidewaystable}[p]
  \centering
  \begin{tabularx}{\textwidth}{@{\extracolsep\fill}c} \hphantom{~} \\ \hphantom{~} \\ \hphantom{~} \\ \end{tabularx}
  \caption{Table of upper limits at 90\% C.L. during X-ray flares and hardness transition states~(TS) analyses. For each source is shown the lifetime, LT, and for each spectra the energy 
  integral, I$(E)$, the differential flux normalisation, $\phi_{0}$, the mean energy flux, $F$ and the fluence, $\mathcal{F_\nu}$.}
  \label{table:ULs}
  \begin{tabularx}{\textwidth}{@{\extracolsep\fill}lccccccccccccc}
    \toprule
      \multirow{2}{*}{Source} & \multirow{2}{*}{LT} &
      \multicolumn{4}{c}{$E^{-2}$} &
      \multicolumn{4}{c}{$E^{-2} \cdot \exp(-E/\text{100\,TeV}$} &
      \multicolumn{4}{c}{$E^{-2} \cdot \exp(-E/\text{10\,TeV})$} \\
    \cmidrule(r){3-6} \cmidrule(r){7-10} \cmidrule(r){11-14}
      & &
      I$(E)$ & $\phi_{0}$ & $F$ & $\mathcal{F_\nu}$ &
      I$(E)$ & $\phi_{0}$ & $F$ & $\mathcal{F_\nu}$ &
      I$(E)$ & $\phi_{0}$ & $F$ & $\mathcal{F_\nu}$ \\
    \midrule
      1A\,0535\texttt{+}262           & 66  & 7.0 & 2.1 & 1.5 & 8.4 & 3.4 & 5.4 & 1.8 & 10  & 2.3 & 1.8 & 4.1 & 24  \\
      1A\,1118\texttt{-}61            & 55  & 6.7 & 2.5 & 1.7 & 8.1 & 3.8 & 4.7 & 1.8 & 8.5 & 2.6 & 1.2 & 3.1 & 15  \\
      4U\,1630\texttt{-}472           & 144 & 7.1 & 1.0 & 0.7 & 9.0 & 3.9 & 1.9 & 0.74& 9.2 & 2.5 & 0.60& 1.5 & 19  \\
      4U\,1636\texttt{-}536           & 43  & 7.0 & 4.2 & 2.9 & 11  & 3.9 & 7.1 & 2.7 & 10  & 2.6 & 1.9 & 4.9 & 18  \\
      4U\,1954\texttt{+}31            & 32  & 7.2 & 3.9 & 2.8 & 7.7 & 3.3 & 12  & 3.8 & 10  & 2.2 & 4.1 & 9.1 & 25  \\
      4U\,\texttt{+}1705\texttt{-}440 & 634 & 7.0 & 0.27& 0.2 & 10  & 3.7 & 0.57& 0.21& 12  & 2.5 & 0.16& 0.41& 23  \\
      Aql\,X\texttt{-}1               & 131 & 7.0 & 1.1 & 0.7 & 8.4 & 3.7 & 2.2 & 0.81& 9.2 & 2.5 & 0.69& 1.7 & 19  \\
      Cir\,X\texttt{-}1               & 232 & 6.8 & 0.89& 0.6 & 12  & 3.8 & 1.6 & 0.62& 12  & 2.5 & 0.48& 1.2 & 24  \\
      Cyg\,X\texttt{-}1               & 228 & 7.0 & 1.2 & 0.8 & 16  & 3.2 & 3.1 & 1.0 & 20  & 2.2 & 0.99& 2.2 & 43  \\
      EX\,O1745\texttt{-}248          & 55  & 7.1 & 2.5 & 1.8 & 8.5 & 3.8 & 4.8 & 1.8 & 8.7 & 2.5 & 1.3 & 3.4 & 16  \\
      Ginga\,1843\texttt{+}00         & 5   & 6.9 & 28  & 19  & 8.3 & 3.6 & 58  & 21  & 9.1 & 2.5 & 14  & 36  & 16  \\
      GRO\,J1008\texttt{-}57          & 146 & 6.8 & 0.95& 0.6 & 8.1 & 3.9 & 1.6 & 0.63& 7.9 & 2.6 & 0.45& 1.2 & 15  \\
      GS\,0834\texttt{-}430           & 74  & 7.2 & 3.0 & 2.1 & 14  & 3.9 & 5.7 & 2.2 & 14  & 2.6 & 1.6 & 4.0 & 25  \\
      GX\,1\texttt{+}4                & 287 & 7.1 & 1.1 & 0.8 & 19  & 3.8 & 2.3 & 0.85& 21  & 2.5 & 0.70& 1.7 & 43  \\
      GX\,304\texttt{-}1              & 200 & 6.7 & 0.86& 0.6 & 10  & 3.9 & 1.5 & 0.58& 9.9 & 2.5 & 0.48& 1.2 & 21  \\
      GX\,339\texttt{-}4              & 184 & 7.1 & 0.45& 0.3 & 5.0 & 3.8 & 0.93& 0.35& 5.6 & 2.5 & 0.25& 0.63& 10  \\
      H\,1417\texttt{-}624            & 64  & 6.6 & 2.6 & 1.7 & 9.3 & 3.8 & 4.7 & 1.8 & 9.7 & 2.5 & 1.3 & 3.2 & 17  \\
      H\,1608\texttt{-}522            & 487 & 7.0 & 0.33& 0.2 & 10  & 3.8 & 0.60& 0.23& 9.7 & 2.5 & 0.19& 0.47& 20  \\
      IGR\,J17091\texttt{-}3624       & 34  & 7.2 & 6.1 & 4.4 & 13  & 3.8 & 12  & 4.7 & 14  & 2.5 & 3.7 & 9.3 & 28  \\
      IGR\,J17464\texttt{-}3213       & 231 & 7.1 & 0.70& 0.5 & 10  & 3.8 & 1.3 & 0.50& 10  & 2.5 & 0.41& 1.0 & 20  \\
      IGR\,J17473\texttt{-}2721         & 6  & 7.2 & 11  & 8.2 & 3.9 & 3.8 & 21  & 8.1 & 3.9 & 2.5 & 6.2 & 15  & 7.5 \\
      \hphantom{SWIFT\,J1910.2\texttt{-}0546 (TS)} & \hphantom{2} & \hphantom{7.1} & \hphantom{170} & \hphantom{120} & \hphantom{18} & \hphantom{3.8} & \hphantom{370} & \hphantom{140} & \hphantom{20} & \hphantom{2.5} & \hphantom{98} & \hphantom{250} & \hphantom{36} \\
      \multicolumn{1}{r}{\hphantom{\multirow{3}{*}{Units:}}} & & & \hphantom{$\cdot10^{-7}$} & \hphantom{$\cdot10^{-6}$} & & & \hphantom{$\cdot10^{-7}$} & \hphantom{$\cdot10^{-6}$} & & & \hphantom{$\cdot10^{-6}$} & \hphantom{$\cdot10^{-6}$} & \\

  \end{tabularx}
\end{sidewaystable}
\begin{sidewaystable}[p]
  \centering
  \begin{tabularx}{\textwidth}{@{\extracolsep\fill}lccccccccccccc}
      \multirow{2}{*}{\hphantom{Source}} & \multirow{2}{*}{\hphantom{LT}} &
      \multicolumn{4}{c}{\hphantom{$E^{-2}$}} &
      \multicolumn{4}{c}{\hphantom{$E^{-2} \cdot \exp(-E/\text{100\,TeV})$}} &
      \multicolumn{4}{c}{\hphantom{$E^{-2} \cdot \exp(-E/\text{10\,TeV})$}} \\
      & &
      \hphantom{I$(E)$} & \hphantom{$\phi_{0}$} & \hphantom{$F$} & \hphantom{$\mathcal{F_\nu}$} &
      \hphantom{I$(E)$} & \hphantom{$\phi_{0}$} & \hphantom{$F$} & \hphantom{$\mathcal{F_\nu}$} &
      \hphantom{I$(E)$} & \hphantom{$\phi_{0}$} & \hphantom{$F$} & \hphantom{$\mathcal{F_\nu}$} \\

      MAXI\,J1409\texttt{-}619          & 24 & 6.7 & 7.1 & 4.7 & 10  & 3.9 & 12  & 4.8 & 9.7 & 2.6 & 3.0 & 7.7 & 16  \\
      MAXI\,J1659\texttt{-}152          & 31 & 7.0 & 4.4 & 3.1 & 8.3 & 3.7 & 9.3 & 3.5 & 9.3 & 2.5 & 3.0 & 7.5 & 20  \\
      MAXI\,J1836\texttt{-}194          & 76 & 7.0 & 1.7 & 1.2 & 8.1 & 3.8 & 3.1 & 1.2 & 7.7 & 2.5 & 0.92& 2.3 & 15  \\
      MXB\,0656\texttt{-}072            & 18 & 6.9 & 6.8 & 4.7 & 7.4 & 3.7 & 13  & 4.7 & 7.3 & 2.5 & 3.9 & 9.5 & 15  \\
      SAX\,J1808.4\texttt{-}3658        & 28 & 7.2 & 6.1 & 4.4 & 11  & 3.8 & 13  & 5.0 & 12  & 2.5 & 3.5 & 8.9 & 22  \\
      SWIFT\,J1539.2\texttt{-}6227      & 23 & 6.6 & 5.4 & 3.6 & 7.1 & 3.9 & 9.2 & 3.6 & 7.0 & 2.6 & 2.4 & 6.1 & 12  \\
      SWIFT\,J1842.5\texttt{-}1124      & 55 & 6.9 & 2.2 & 1.5 & 7.2 & 3.8 & 4.0 & 1.5 & 7.2 & 2.5 & 1.2 & 3.0 & 14  \\
      SWIFT\,J1910.2\texttt{-}0546      & 73 & 7.0 & 1.5 & 1.1 & 6.7 & 3.8 & 2.9 & 1.1 & 6.9 & 2.5 & 0.86& 2.2 & 14  \\
      XTE\,J1652\texttt{-}453           & 40 & 7.2 & 5.7 & 4.1 & 14  & 3.8 & 9.8 & 3.7 & 13  & 2.6 & 2.8 & 7.3 & 25  \\
      XTE\,J1752\texttt{-}223           & 78 & 7.1 & 2.0 & 1.4 & 10  & 3.8 & 4.4 & 1.7 & 11  & 2.5 & 1.2 & 3.0 & 20  \\
      XTE\,J1810\texttt{-}189           & 5  & 7.1 & 30  & 21  & 10  & 3.9 & 51  & 20  & 8.9 & 2.6 & 15  & 39  & 17  \\
      XTE\,J1946\texttt{+}274           & 73 & 7.0 & 1.7 & 1.2 & 7.7 & 3.4 & 4.1 & 1.4 & 8.7 & 2.2 & 1.6 & 3.5 & 22  \\
    \midrule
      GX\,339\texttt{-}4 (TS)           & 3  & 7.2 & 100 & 75  & 18  & 4.0 & 190 & 75  & 18  & 2.6 & 53  & 140 & 33  \\
      H\,1608\texttt{-}522 (TS)         & 10 & 7.1 & 27  & 19  & 17  & 3.9 & 58  & 23  & 20  & 2.6 & 18  & 48  & 42  \\
      IGRJ\,17091\texttt{-}3624 (TS)    & 2  & 7.3 & 160 & 120 & 23  & 3.9 & 330 & 130 & 25  & 2.6 & 92  & 240 & 46  \\
      IGRJ\,17464\texttt{-}3213 (TS)    & 8  & 7.1 & 19  & 13  & 9.4 & 3.8 & 32  & 12  & 8.7 & 2.6 & 8.2 & 21  & 15  \\
      MAXI\,J1659\texttt{-}152 (TS)     & 4  & 7.1 & 56  & 40  & 15  & 3.8 & 120 & 44  & 17  & 2.5 & 36  & 91  & 35  \\
      SWIFT\,J1910.2\texttt{-}0546 (TS) & 2  & 7.1 & 170 & 120 & 18  & 3.8 & 370 & 140 & 20  & 2.5 & 98  & 250 & 36  \\
      XTE\,J1652\texttt{-}453 (TS)      & 49 & 7.1 & 3.8 & 2.7 & 11  & 3.8 & 8.0 & 3.1 & 13  & 2.5 & 2.4 & 6.2 & 26  \\
      XTE\,J1752\texttt{-}223 (TS)      & 3  & 7.0 & 140 & 97  & 22  & 3.8 & 270 & 100 & 23  & 2.6 & 68  & 170 & 39  \\
    \midrule
      \multicolumn{1}{r}{\multirow{3}{*}{Units:}} & & & $\cdot10^{-7}$ & $\cdot10^{-6}$ & & & $\cdot10^{-7}$ & $\cdot10^{-6}$ & & & $\cdot10^{-6}$ & $\cdot10^{-6}$ & \\
      & \multicolumn{13}{c}{
        \begin{scriptsize}$\left[\text{LT}\right] = \text{days}$\end{scriptsize} ~ $|$ ~
        \begin{scriptsize}$\left[\text{I}(E)\right] = \text{GeV}^{2}$\end{scriptsize}
      } \\
      & \multicolumn{13}{c}{
        \begin{scriptsize}$\left[\phi_{0}\right] = \text{GeV}^{-1}\,\text{cm}^{-2}\,\text{s}^{-1}$\end{scriptsize} ~ $|$ ~
        \begin{scriptsize}$\left[F\right] = \text{GeV}\,\text{cm}^{-2}\,\text{s}^{-1}$\end{scriptsize} ~ $|$ ~
        \begin{scriptsize}$\left[\mathcal{F}\right] = \text{GeV}\,\text{cm}^{-2}$\end{scriptsize}
      } \\
    \bottomrule
  \end{tabularx}
  \vspace*{1pt}
  \begin{tabularx}{\textwidth}{@{\extracolsep\fill}c} \hphantom{~} \\ \hphantom{~} \\ \hphantom{~} \\ \hphantom{~} \\ \hphantom{~} \\ \end{tabularx}
\end{sidewaystable}
}

In the following, the fluence upper limits derived above are discussed in the light of theoretical models of neutrino production in X-ray binaries. Microquasars are considered first, followed by X-ray binaries that do not exhibit relativistic jets.\\

\subsection{Discussion on the microquasar case}\label{sub:muqso}

Neutrinos can be produced in microquasars either in p-$\gamma$ \cite{bib:Distefano2002, bib:Romero2008} or p-p interactions \cite{bib:Romero2003, bib:Christiansen2006}, as discussed in the introduction. However, both p-$\gamma$ and p-p models usually assume that the magnetic energy is in equipartition with the kinetic energy in the jets, leading to large magnetic field intensities. In this context, Reynoso et al~\cite{bib:Reynoso2009} find that pions and muons produced close to the compact object can be strongly affected by synchrotron losses, decreasing significantly the expected high-energy neutrino flux. 
However, in high-mass microquasars, where the companion star present a strong wind, the interaction of energetic protons in the jet with matter of dense clumps of the wind could produce detectable neutrino fluxes \cite{bib:Reynoso2009}, but still below the current sensitivity of ANTARES (see figure 11 of \cite{bib:Reynoso2009}).

Furthermore, Bednarek~\cite{bib:Bednarek2005} considers a binary system composed of a compact object orbiting a Wolf-Rayet star, in which nuclei accelerated in the jet efficiently lose neutrons as a result of photo-disintegration process. Consequently, they claims that these neutrons can produce neutrinos in collisions with matter from the accretion disk and/or the massive companion star\cite{bib:Bednarek2005}. However, this model principally applies to Cyg~X-3 which is not covered in the present analysis. 

\begin{figure}[ht!]
\centering
\includegraphics[width=\textwidth]{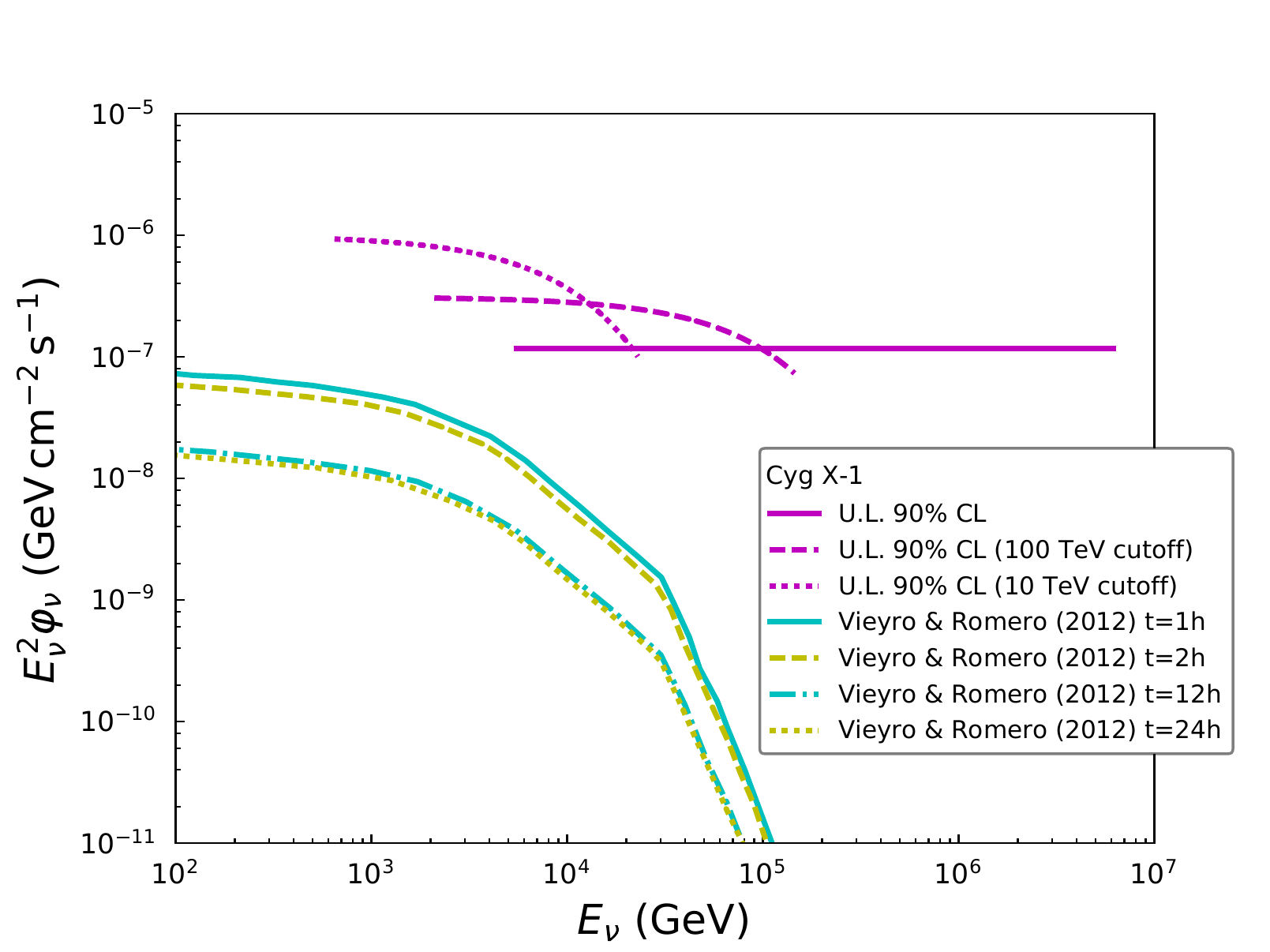}
\caption{Upper limits at 90\%~C.L. on the neutrino energy flux obtained in this analysis in 
the case of $E^{-2}$, $E^{-2}\exp(-E/100~\rm{TeV})$ and $E^{-2}\exp(-E/10~\rm{TeV})$ neutrino energy spectra compared with the 
expectations~\cite{bib:Vieyro2012} for the microquasar Cyg~X-1 at different times after the rise of the neutrino flare.
}
\label{fig:Vieyro}
\end{figure}

Recently, it was shown that the non-thermal emission of Cyg~X-1 can be explained by a static corona model which is supported by magnetic pressure, and does not imply the presence of a relativistic outflow~\cite{bib:Romero2010}. Upon this model, the neutrino emission driven by the injection of relativistic particles into that strongly magnetized corona was studied \cite{bib:Vieyro2012}. Their predictions are shown in Figure \ref{fig:Vieyro} and compared to the ANTARES upper limits. Unfortunately, the location of Cyg~X-1 in the Northern hemisphere, does not enable to further constrain this model yet.

U.L. imposed by ANTARES allow to constrain the most favorable hadronic models for microquasars. Hereafter, the neutrino energy flux predictions, $F_{\mathrm{pred}}$, has been computed for seven microquasars according to the photohadronic model ~\cite{bib:Distefano2002}, based on a previous work Levinson \& Waxman \cite{bib:Levinson2001}. Since they are less stringent, the energy flux 
upper limits of the transition states are not discussed in the following. Thus, in the following, only the neutrino energy flux upper limits 
related to the hard state periods are considered. Using the latest measurements of their distance and of the jet parameters, the model predicts the neutrino energy flux based on the radio 
luminosity of the jets observed in radio during flares. The derived neutrino energy flux depends on the fraction of jet kinetic energy, $L_{\mathrm{jet}}$, converted respectively 
to relativistic electrons and magnetic field, $\eta_\mathrm{e}$, and protons $\eta_\mathrm{p}$, and the fraction of proton energy converted into pions, $f_\pi$, which depends in turn on 
the energy to which protons are accelerated. 
Resolved and unresolved sources are considered separately. In the following, resolved sources refer to microquasar jets resolved by radio interferometry which enables the physical parameters 
of the jet to be probed (Table \ref{table:compareDistefano1}).\\  

For resolved sources, the neutrino energy flux is estimated from the radio flux density, $S_f$, at frequency $f$, the distance of the source, $d$, the size of the emitting 
region, $l$, the jet Lorentz factor, $\Gamma$, the jet velocity, $\beta$, the angle, $\theta$, between the 
line of sight and the jet, and the jet opening angle, $\psi$. The ratio of the minimum and maximum electron Lorentz factors, respectively $\gamma_{\mathrm{min}}$ and  
$\gamma_{\mathrm{max}}$, is assumed to be equal to 100, while $\psi$ is taken equal to a conservative value of 20$^\circ$ except in the case of Cyg X-1~\cite{bib:MillerJones2006}. All the 
other parameters have been retrieved from the literature and are listed in Table~\ref{table:compareDistefano1} along with 
their uncertainties and references. When $\theta$ is not observationally constrained, all the values between 0 and 90$^\circ$ are considered. Similarly, if the bulk Lorentz 
factor, $\Gamma$, of the jet is poorly known, all the potential values are tested. These uncertainties, together with the error on the other jet parameters, 
are taken into account to derive the range of neutrino energy fluxes $F_{\mathrm{pred}}$ satisfying the model which is linearly dependent on $\eta_\mathrm{e}/\eta_\mathrm{p}$.\\

\begin{sidewaystable}[p]
\small
  \centering
\caption{List of jet parameters and neutrino energy flux expected~\cite{bib:Distefano2002} for resolved microquasars, upper limits at 90\% C.L. on the neutrino energy fluxes given 
by ANTARES ($F$) for a $E^{-2}\mathrm{exp}(-\sqrt{E/100\mathrm{\ TeV}})$ neutrino energy spectra, and upper limits on $\eta_\mathrm{p}/\eta_\mathrm{e}$ resulting from the ANTARES upper limits 
compared to the expectations~\cite{bib:Distefano2002}.}
\label{table:compareDistefano1}
\centering
\begin{tabular}{|c||c|c|c|c|c|c|c|c|c|c|c|}
\hline Source  & $d$ & $f$ & $S_f$ & $l$ & $\psi$ & $\theta$  & $\Gamma$ & $\eta_\mathrm{e}/\eta_\mathrm{p}F_\mathrm{pred}$ & $F$  U.L.&$\eta_\mathrm{p}/\eta_\mathrm{e}$ U.L. & Ref.  \\
& (kpc) & (GHz) & (mJy) & ($10^{15}$ cm) & (deg) & (deg) & & (erg~s$^{-1}$~cm$^{-2}$) & (erg~s$^{-1}$~cm$^{-2}$) &  &\\
\hline
Cyg X-1 & $1.8^{+0.56}_{-0.56}$ & 8.4 & 3.1 & 0.404 & < 2 &  20 -- 70&  > 1.6 & 8.0 10$^{-12}$ -- 1.3 10$^{-10}$ & 1.6 10$^{-9}$& 200 & \cite{bib:Stirling2001} \\
Cir X-1 & 7.8 - 10.5 & 8.4 & 104 & 2.3 & < 20 & 0 -- 90 & > 1 & 1.8 10$^{-11}$ -- 1.9 10$^{-8}$ & 1.0 10$^{-9}$& 56 & \cite{bib:MillerJones2012} \\
Cir X-1 & $9.4^{+0.8}_{-1.0}$ & 8.4 & 104 & 2.8  & < 20 & < 3 & > 22 & 8.2 10$^{-9}$ -- 1.0 10$^{-5}$ & 1.0 10$^{-9}$ &  0.1 & \cite{bib:Heinz2015} \\
H1743-322 & $8.5^{+0.8}_{-0.8}$ & 8.4 & 6.6 & 1.72 & < 20 & $75^{+3}_{-3}$ & 1.02 -- 1.63 & 1.6 10$^{-12}$ -- 2.2 10$^{-12}$ &  8.5 10$^{-10}$ & 531 & \cite{bib:Corbel2005,bib:Steiner2012} \\
MAXI J1659-152 & $7^{+3}_{-3}$ & 5.0 & 7.3 & 0.31 & < 20 & 0 -- 90 & > 1 & 6.3 10$^{-13}$ -- 8.2 10$^{-9}$ & 5.2 10$^{-9}$ & 8254  & \cite{bib:Paragi2013} \\
\hline
\end{tabular}
\end{sidewaystable}

\begin{figure}[ht!]
\centering
\includegraphics[width=0.47\textwidth]{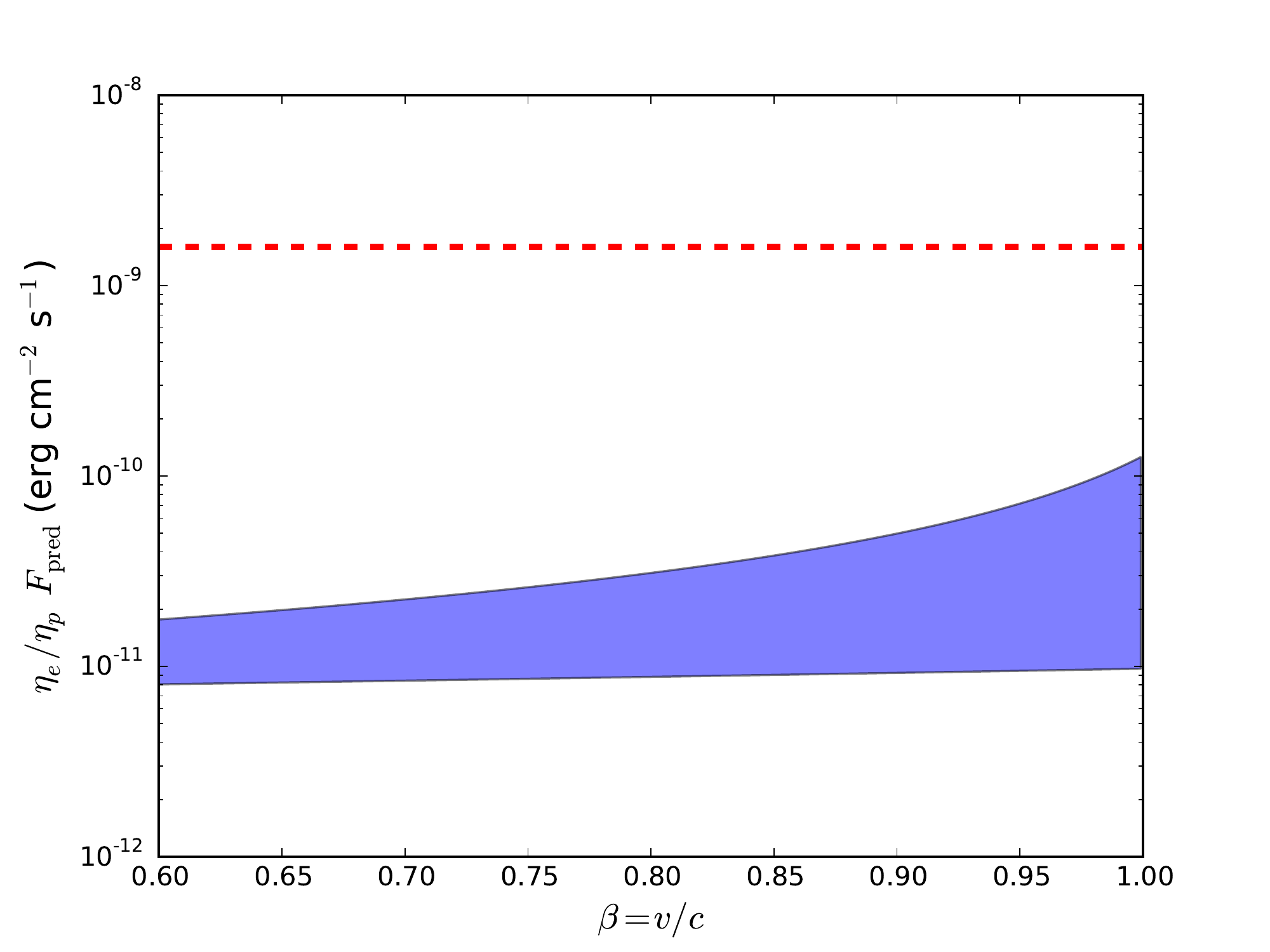}
\includegraphics[width=0.47\textwidth]{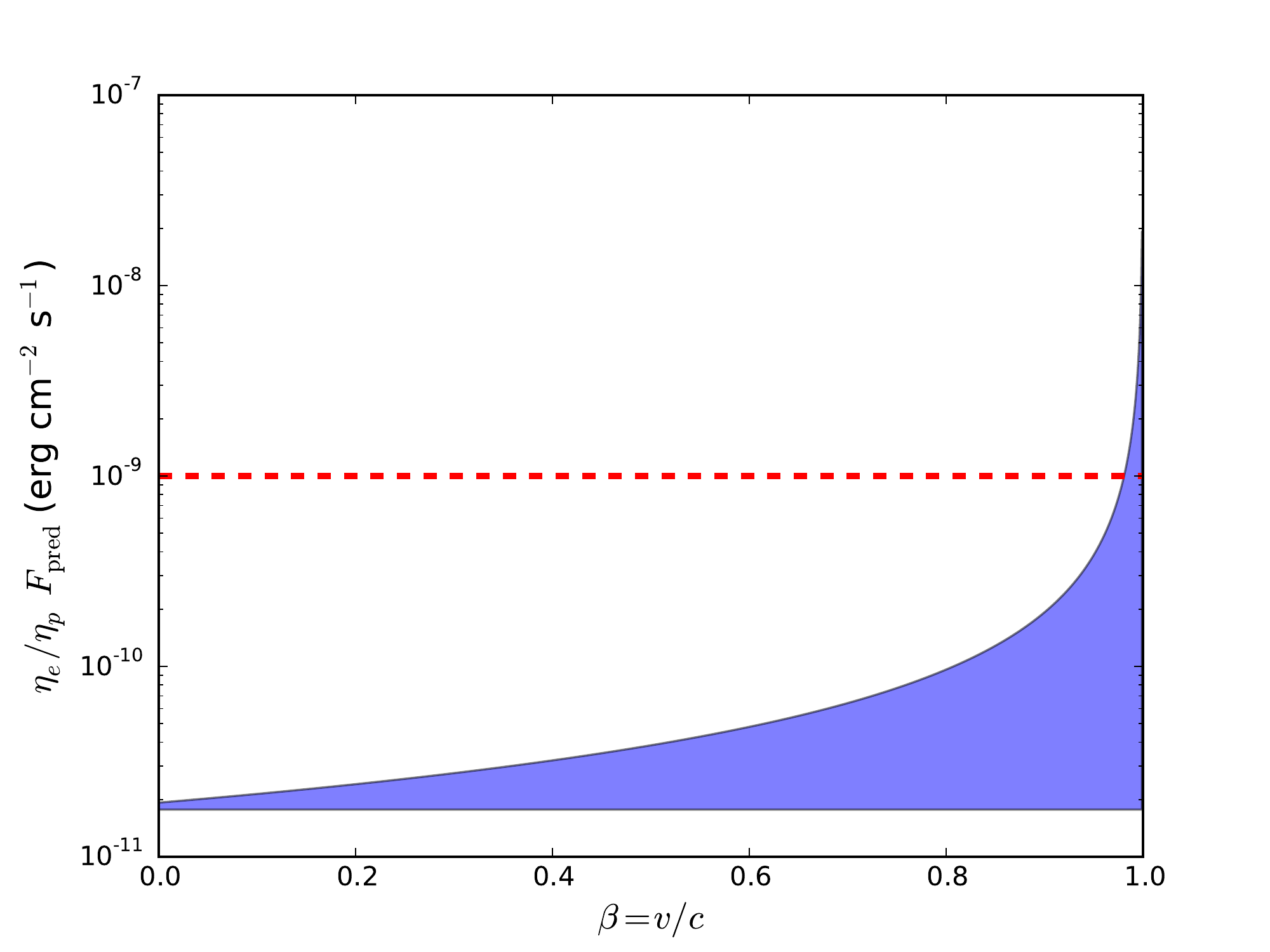}
\includegraphics[width=0.47\textwidth]{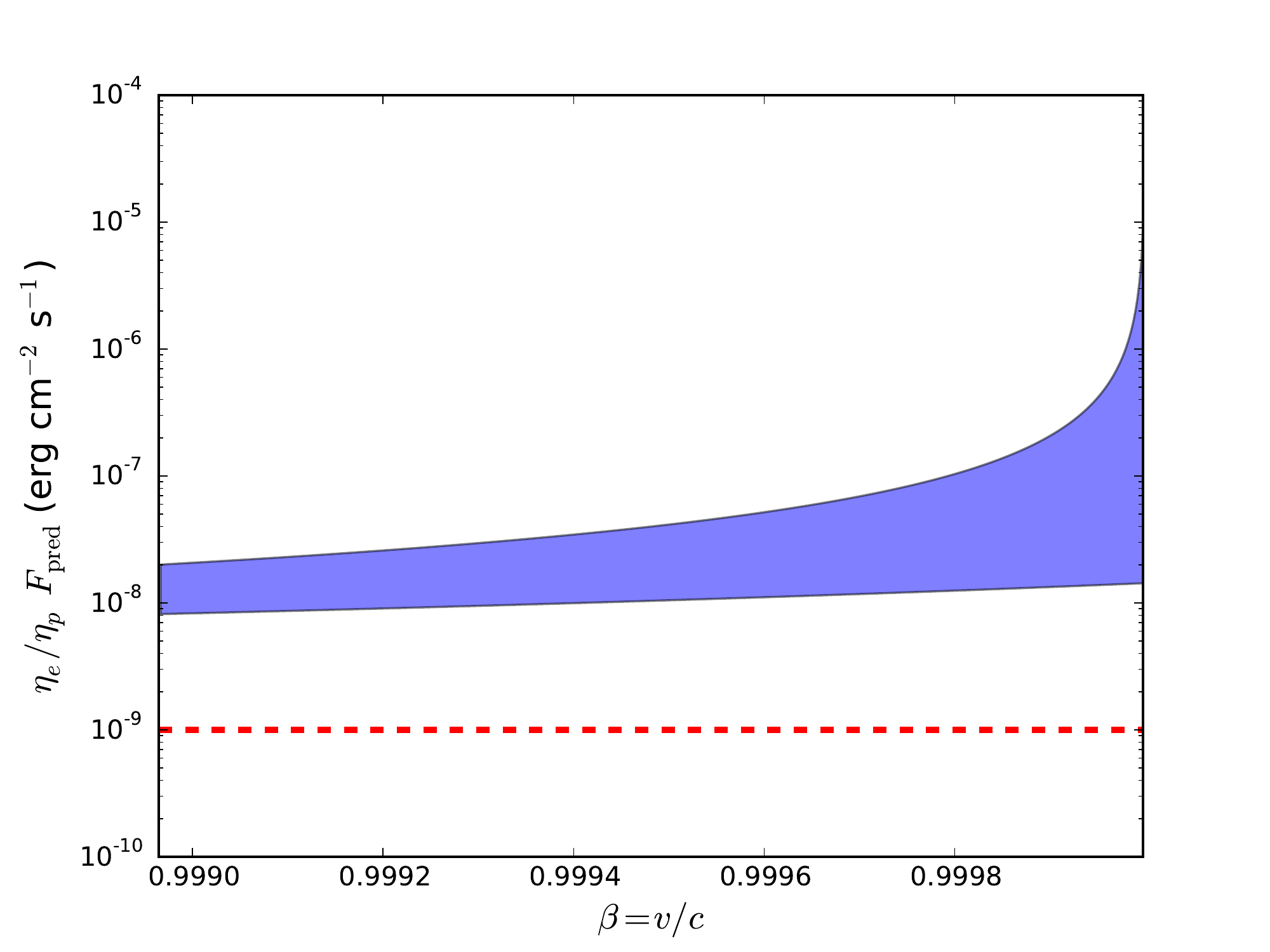}
\includegraphics[width=0.47\textwidth]{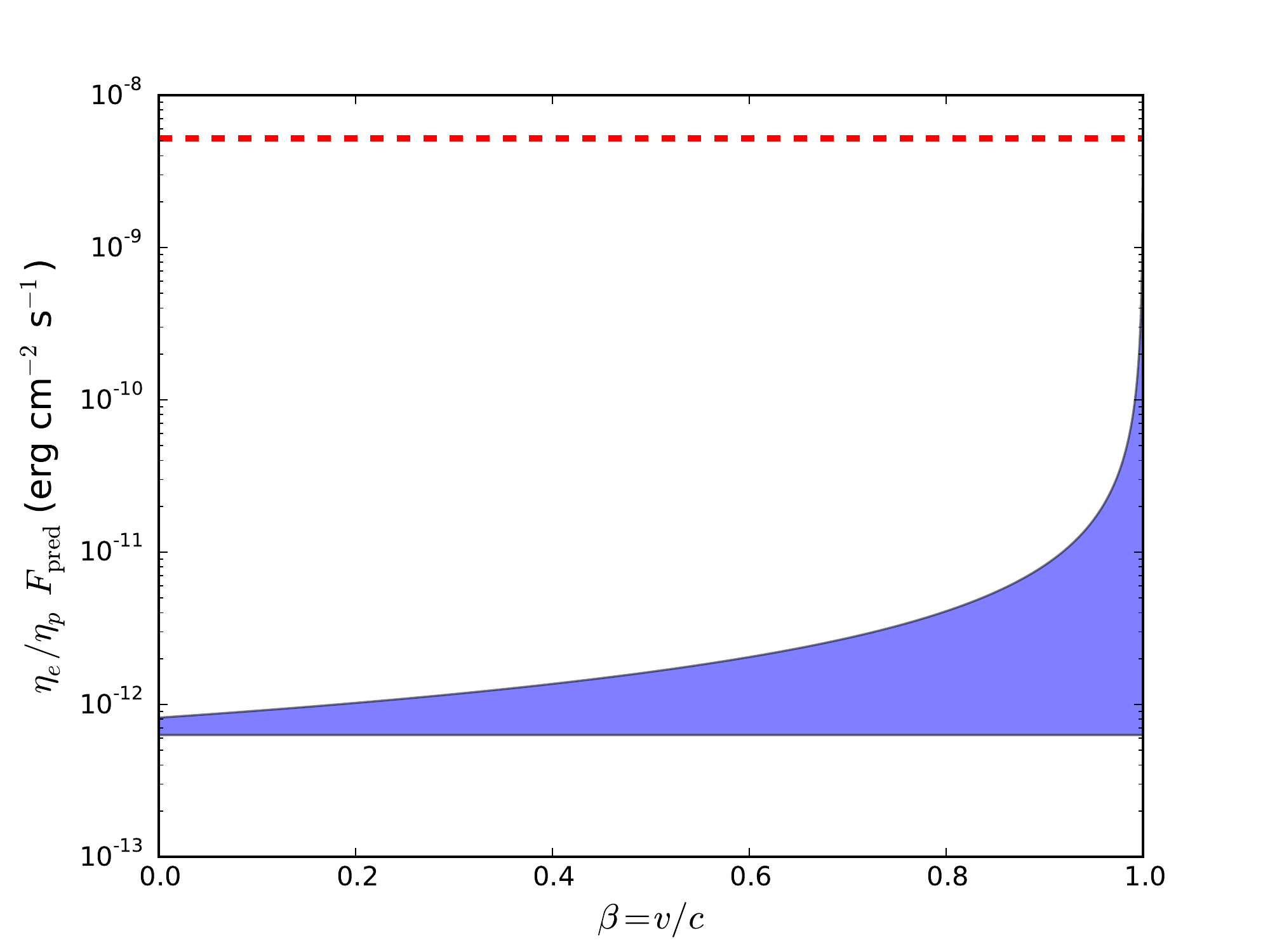}
\caption{Comparison of the energy flux upper limit at 90\% C.L. provided by ANTARES (red dashed line) with the predictions~\cite{bib:Distefano2002} as a function of the jet 
velocity, $\beta$: Cyg X-1 (top left), Cir X-1 with jet parameters~\cite{bib:MillerJones2012} (top right), Cir X-1 with jet parameters~\cite{bib:Heinz2015} (bottom left) and 
MAXI J1659-152 (bottom right).}
\label{fig:CygX-1}
\end{figure}

The resulting predictions are compared with the upper limits obtained with ANTARES data under the hypothesis of a cutoff at 100 TeV in the neutrino flux, to account for limits in the 
acceleration process included in the model~\cite{bib:Distefano2002}. As an example, Fig.~\ref{fig:CygX-1} shows how the predicted flux compares with this result as a function of the 
jet parameter $\beta$ for resolved microquasars Cyg X-1, Cir X-1 and MAXI J1659-152~\footnotemark[5]. Comparing the predicted flux and the ANTARES neutrino energy flux upper limits, upper limits at 90\% C.L. on $\eta_\mathrm{p}/\eta_\mathrm{e}$ are set. These limits have been derived taking into account the discrepancy between Lorentz 
factors reported in radio observations, and uncertainties on the opening angle of the jet, the distance of the source, and on the inclination angle between the line of sight and the jet. 
Results are given for Cyg X-1, Cir X-1 and MAXI J1659-152 in Fig.~\ref{fig:CygX-1} and in Table~\ref{table:compareDistefano1}.

\footnotetext[5]{The energy flux upper limit obtained for H 1743-322 is around 2 orders of magnitude 
higher than the expectations~\cite{bib:Distefano2002}. Thus, this source is not included in Fig.~\ref{fig:CygX-1}.}

However, the potential variability of the Lorentz factor during a burst and between the periods of activity of the source are not taken into account in this calculation. Thus, constraints on baryon loading may have different implications: the proton component in the jet can be negligible in comparison with the electromagnetic component, the proton energy fraction converted to pions can be less significant than the values considered~\cite{bib:Distefano2002}, 
and/or the jet Lorentz factor is lower than the constraints set by radio observations. 

\begin{figure}[ht!]
\centering
\includegraphics[width=\textwidth]{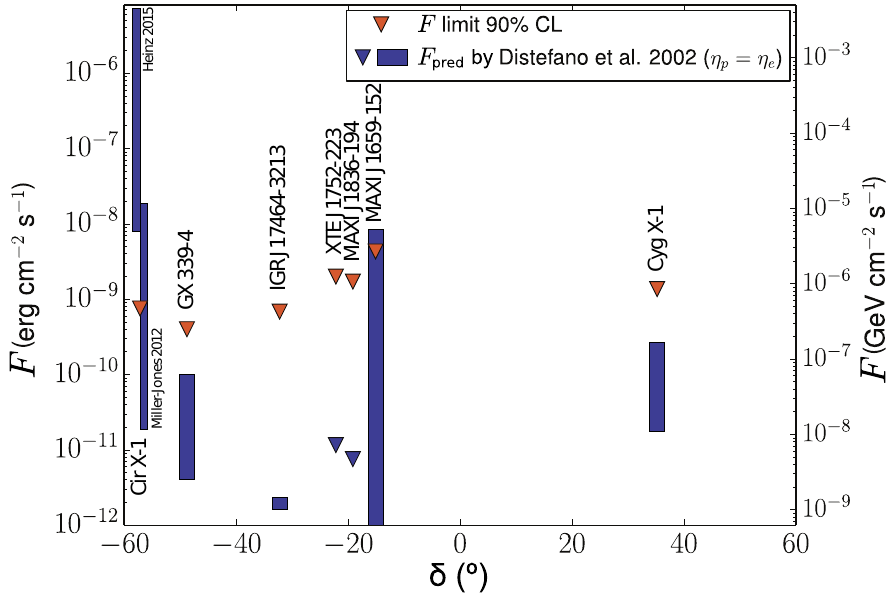}
\caption{Upper limits at 90\%~C.L. on the neutrino energy flux obtained in this analysis considering a $E^{-2}\exp(-\sqrt(E/100~\rm{TeV}))$ spectra, compared with the 
expectations~\cite{bib:Distefano2002} assuming energy equipartition between electrons and protons. The blue rectangles show the expectations from~\cite{bib:Distefano2002} 
taking into account the uncertainty on the jet velocity. For the unresolved microquasars XTE J1752-223 and MAXI J1836-194, a single energy flux prediction value is given, since 
no detail on the jet velocity is available in the literature.
}
\label{fig:DiStefano}
\end{figure}

For unresolved sources, the jet kinetic power is evaluated from the jet synchrotron luminosity derived from the flux density, $S_{f_{\mathrm{break}}}$, at 
the frequency break, $f_{\mathrm{break}}$, between optically thick and optically thin radio emission, and the spectral index, $\alpha_R$~\cite{bib:Distefano2002}. These values are reported in 
Table~\ref{table:compareDistefano2}. When no spectral index value is provided in the literature, $\alpha_R=0$ is assumed as given by the standard jet radio emission theory \cite{bib:Blandford1979}. Again, 
the neutrino energy flux is linearly dependent on $f_\pi$ and $\eta_\mathrm{e}/\eta_\mathrm{p}$. The predicted neutrino 
energy flux $\eta_\mathrm{e}/\eta_\mathrm{p}F_{\mathrm{pred}}$ and the upper limits on $\eta_\mathrm{p}/\eta_\mathrm{e}$ are given in Table~\ref{table:compareDistefano2}. The results for 
both resolved and unresolved sources are summarised in Figure \ref{fig:DiStefano}.\\

\begin{sidewaystable}[p]
  \centering
\caption{List of jet parameters, $\alpha_R$, $f_{\mathrm{break}}$ and $S_{f_{\mathrm{break}}}$, neutrino flux energy expected~\cite{bib:Distefano2002} for unresolved microquasars ($\eta_\mathrm{e}/\eta_\mathrm{p}F_\mathrm{pred}$), 
upper limits on the neutrino energy flux given by ANTARES ($F$) for $E^{-2}\mathrm{exp}(-\sqrt{E/100\mathrm{\ TeV}})$ neutrino energy spectra and upper limits on 
$\eta_\mathrm{p}/\eta_\mathrm{e}$. For GX 339-4, two sets of values, related to two observing periods, are given~\cite{bib:Russell2013}. For XTE 1752-223 and MAXI J1836-194, the upper 
limits on $\eta_\mathrm{p}/\eta_\mathrm{e}$ are computed assuming $\eta_\mathrm{e}/\eta_\mathrm{p}f_\nu$ equal to 1.35 10$^{-11}$ and 9.11 10$^{-12}$ respectively.}
\label{table:compareDistefano2}
\centering
\begin{tabular}{|c||c|c|c|c|c|c|}
\hline Source name &  $\alpha_R$ & log($f_{\mathrm{break}}$) & $S_{f_{\mathrm{break}}}$ & $\eta_\mathrm{e}/\eta_\mathrm{p}F_\mathrm{pred}$ & $F$ U.L. & $\eta_\mathrm{p}/\eta_\mathrm{e}$ U.L.  \\
&&log(GHz)&(mJy)&(erg s$^{-1}$ cm$^{-2}$)&(erg s$^{-1}$ cm$^{-2}$) & \\
\hline
GX 339-4  & 0.08  & 14.26$^{+0.12}_{-0.12}$ & 23$^{+18}_{-18}$  & 3.99 10$^{-12}$  & 5.0 10$^{-10}$ & 125  \\
GX 339-4  & 0.29 & 13.67$^{+0.25}_{-0.25}$& 251$^{+193}_{-193}$ & 1.07 10$^{-10}$ & 5.0 10$^{-10}$ & 5 \\
XTE 1752-223 & 0 & < 14.26 & 6.14$^{+5.77}_{-5.77}$ & < 1.35 10$^{-11}$ & 2.4 10$^{-9}$ & 178 \\
MAXI J1836-194  & 0 & < 13.40 & 57$^{+1}_{-1}$ & < 9.11 10$^{-12}$ & 2.1 10$^{-9}$ & 231 \\

\hline
\end{tabular}
\end{sidewaystable}

\begin{figure}[ht!]
\centering
\includegraphics[width=\textwidth]{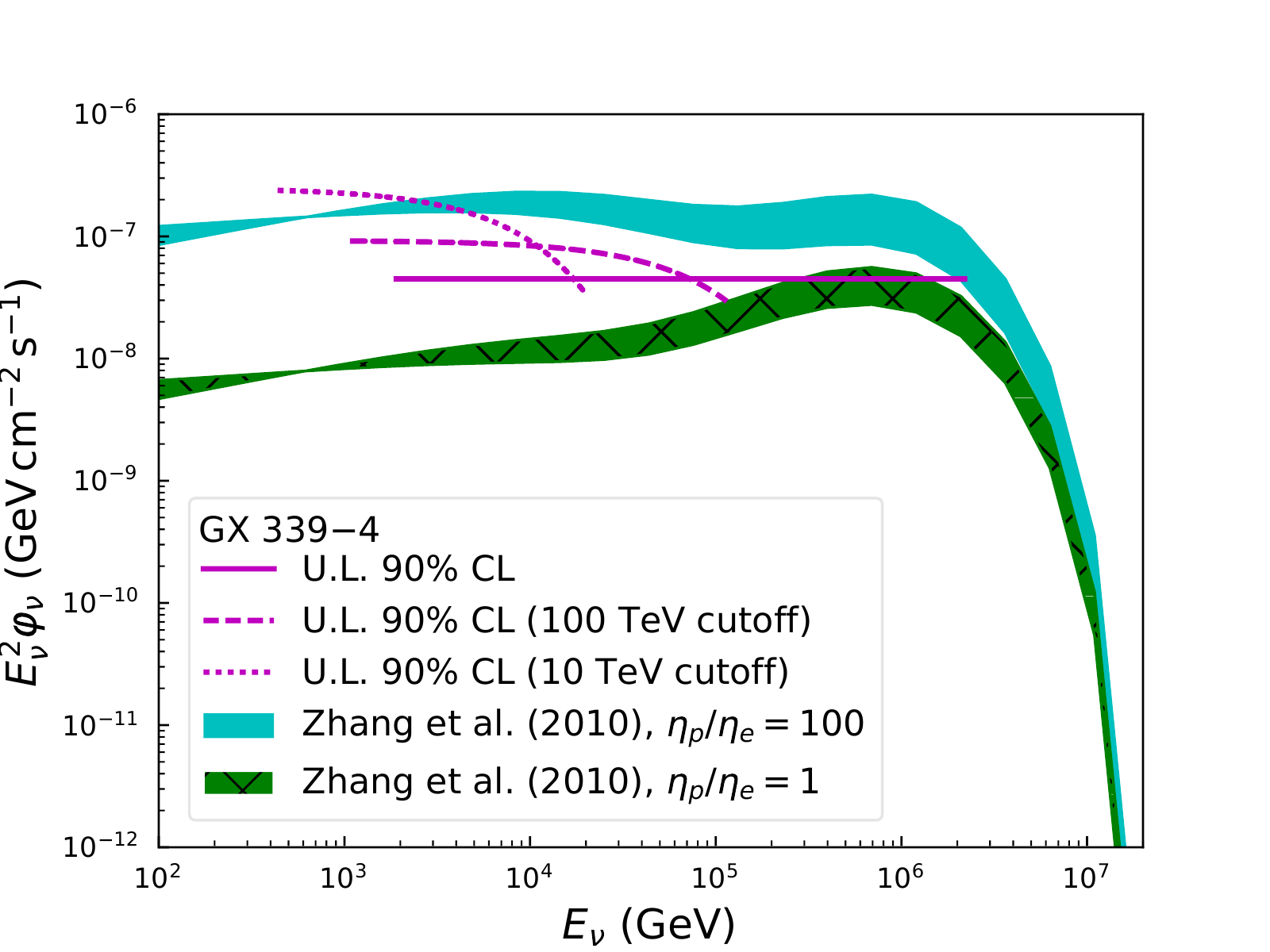}
\caption{Upper limits at 90\% C.L. on the neutrino flux for GX\,339\texttt{-}4, with the energy spectrum described in Section 4, compared to the predictions~\cite{bib:Zhang2010} for 
spectral indexes of the injected particles $-1.8>\alpha>-2.0$ and the ratio $n_{\mathrm{p}}$/$n_{\mathrm{e}}$ equal to 1 and 100.}
\label{fig:Results_model1}
\end{figure}

In \cite{bib:Zhang2010}, the authors have provided a calculation of the high-energy neutrino emission from GX\,339\texttt{-}4 in the hypothesis that the primary spectrum of the 
injected particles in the jets has spectral indexes $-1.8>\alpha>-2.0$ and that the ratio between proton and electron energy is equal to 1~and~100 (Figure~\ref{fig:Results_model1}). The 
model with a ratio $\eta_{\mathrm{p}}$/$\eta_{\mathrm{e}}$ equal to 100 is excluded by the present limit.\\

\subsection{Discussion on the case of X-ray binaries without relativistic jet}

\begin{figure}[ht!]
\centering
\includegraphics[width=\textwidth]{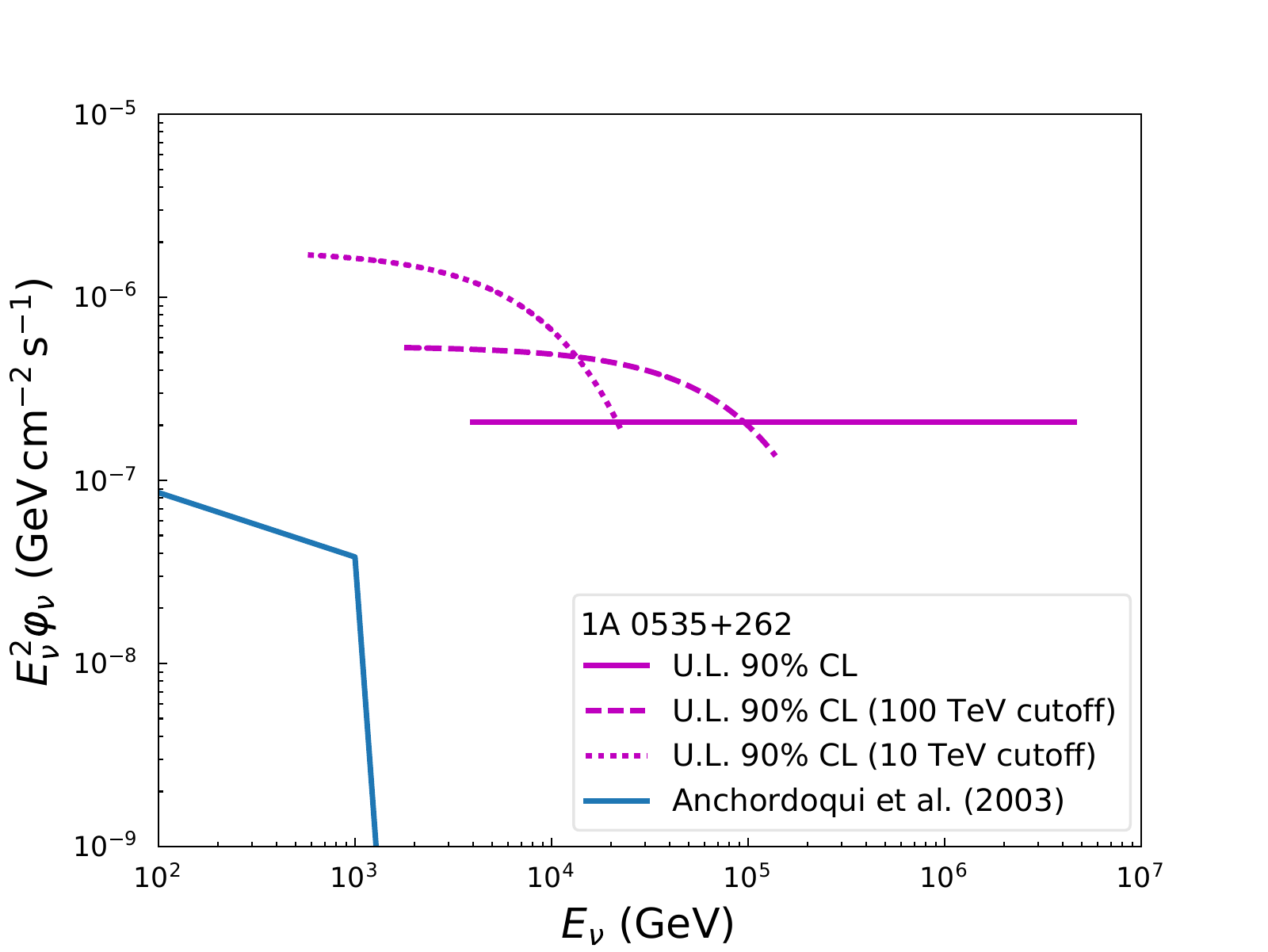}
\caption{Upper limits at 90\%~C.L. on the neutrino energy flux obtained in this analysis in 
the case of $E^{-2}$, $E^{-2}\exp(-E/100~\rm{TeV})$ and $E^{-2}\exp(-E/10~\rm{TeV})$ neutrino energy spectra compared with the 
expectations~\cite{bib:Anchordoqui2003} for the X-ray binary 1A 0535+262.
}
\label{fig:Anchordoqui}
\end{figure}

As stated in Section \ref{TSperiods}, a large part of the known sample of galactic X-ray binaries does not exhibit relativistic jets but neutrino production may take place though. For example, significant potential drops can develop into the magnetosphere of an accreting neutron star as the one hosted by 1A 0535+26. Protons, accelerated in these gaps to energies greater than 100 TeV, can impact onto the accretion disk, finally producing high-energy neutrinos under specific conditions of disk density \cite{bib:Anchordoqui2003}. As seen in Figure \ref{fig:Anchordoqui}, the current upper limits do not enable to challenge this model which predicts a very low neutrino flux above 1 TeV.

Furthermore, Bednarek~\cite{bib:Bednarek2009} has considered a photo-hadronic emission model in which hadrons are accelerated within the inner pulsar rotating magnetosphere. Hadrons consequently produce gamma-rays and neutrinos in collision with thermal radiation from hot spots on the neutron star surface. Although ANTARES cannot confirm this model yet, it might be constrained by kilometer-scale detectors such as KM3NeT~\cite{bib:KM3NET}.

\section{Conclusion}
This paper discusses the time-dependent search for neutrinos from X-ray binaries using the data taken with the full ANTARES detector between 2008 and 2012. This 
search has been applied to a list of 33 XRB sources, 8 of them during hardness transition periods. The search did not result in a statistically significant excess above 
the expected background from atmospheric neutrino and muon events. The most significant correlation during X-ray flares is found for the source GX\,1\texttt{+}4, for which 
3 neutrino candidate events were detected in time/spatial coincidence with X-ray emission. However, the post-trial probability is 72\%, thus compatible with background fluctuations. A 
comparison with predictions from several models shows that for some sources, the upper limits start to constrain the parameter space of the expectations from hadronic jet 
emission models. Therefore, with additional data from ANTARES and with the order of magnitude sensitivity improvement expected from the next generation neutrino telescope, 
KM3NeT~\cite{bib:KM3NET}, the prospects for future searches for neutrino emission from X-ray binaries are very promising. 


\acknowledgments

The authors acknowledge the financial support of the funding agencies:
Centre National de la Recherche Scientifique (CNRS), Commissariat \`a
l'\'ener\-gie atomique et aux \'energies alternatives (CEA),
Commission Europ\'eenne (FEDER fund and Marie Curie Program),
Institut Universitaire de France (IUF), IdEx program and UnivEarthS
Labex program at Sorbonne Paris Cit\'e (ANR-10-LABX-0023 and
ANR-11-IDEX-0005-02), Labex OCEVU (ANR-11-LABX-0060) and the
A*MIDEX project (ANR-11-IDEX-0001-02),
R\'egion \^Ile-de-France (DIM-ACAV), R\'egion
Alsace (contrat CPER), R\'egion Provence-Alpes-C\^ote d'Azur,
D\'e\-par\-tement du Var and Ville de La
Seyne-sur-Mer, France; Bundesministerium f\"ur Bildung und Forschung
(BMBF), Germany; Istituto Nazionale di Fisica Nucleare (INFN), Italy;
Stichting voor Fundamenteel Onderzoek der Materie (FOM), Nederlandse
organisatie voor Wetenschappelijk Onderzoek (NWO), the Netherlands;
Council of the President of the Russian Federation for young
scientists and leading scientific schools supporting grants, Russia;
National Authority for Scientific Research (ANCS), Romania; 
Mi\-nis\-te\-rio de Econom\'{\i}a y Competitividad (MINECO):
Plan Estatal de Investigaci\'{o}n (refs. FPA2015-65150-C3-1-P, -2-P and
-3-P, (MINECO/FEDER)), Severo Ochoa Centre of Excellence and MultiDark
Consolider (MINECO), and Prometeo and Grisol\'{i}a programs (Generalitat
Valenciana), Spain; Agence de  l'Oriental and CNRST, Morocco. We also
 acknowledge the technical support of Ifremer, AIM and Foselev Marine
 for the sea operation and the CC-IN2P3 for the computing facilities.

\end{document}